\documentclass[seceq]{ptptex}
\usepackage[dvips]{graphicx}
\usepackage{wrapft}
\usepackage{bm}
\usepackage{mathrsfs}

\def\lesim{\m@thcombine<\sim}
\def\gesim{\m@thcombine>\sim}
\def\lessgtr{\m@thcombine<>}
\def\gtrless{\m@thcombine><}

\newcommand{\bra}[1]{\left\langle #1 \right|}
\newcommand{\brared}[1]{\langle #1 ||}

\newcommand{\product}[2]{\left\langle #1 | #2 \right\rangle}

\newcommand{\ket}[1]{\left| #1 \right\rangle}
\newcommand{\ketred}[1]{|| #1 \rangle}

\newcommand{\adag}{a^{\dagger}}

\newcommand{\Ahat}{\hat{A}}

\newcommand{\Atdag}{\hat{A}^{(\tau)\dagger}}

\newcommand{\Bhat}{\hat{B}}

\newcommand{\Btdag}{\hat{B}^{(\tau)\dagger}}

\newcommand{\cdag}{c^{\dagger}}

\newcommand{\Hhat}{\hat{H}}

\newcommand{\Ihat}{\hat{I}}

\newcommand{\hhat}{\hat{h}}
\newcommand{\fp}{f^{(+)}}

\newcommand{\fm}{f^{(-)}}
\newcommand{\Fhat}{\hat{F}}

\newcommand{\Fhatp}{\hat{F}^{(+)}}
\newcommand{\Fhatm}{\hat{F}^{(-)}}
\newcommand{\Fhatpm}{\hat{F}^{(\pm)}}

\newcommand{\Hc}{{\cal H}}

\newcommand{\Jc}{{\cal J}}

\newcommand{\Ec}{{\cal E}}

\newcommand{\ddg}{d^{\dagger}}

\newcommand{\Nhat}{\hat{N}}
\newcommand{\Nt}{\tilde{N}}

\newcommand{\Dc}{{\mathscr D}}
\newcommand{\Ddag}{\hat{D}^{\dagger}}

\newcommand{\Dhat}{\hat{D}}
\newcommand{\Ghat}{\hat{G}}

\newcommand{\Qhat}{\hat{Q}}
\newcommand{\Rhat}{\hat{R}}
\newcommand{\Phat}{\hat{P}}

\newcommand{\Psihat}{\hat{\Psi}}
\newcommand{\Qdag}{Q^{\dagger}}
\newcommand{\That}{\hat{\Theta}}

\newcommand{\Fp}{F^{(+)}}
\newcommand{\Fm}{F^{(-)}}
\newcommand{\Rp}{R^{(+)}}
\newcommand{\Rm}{R^{(-)}}

\newcommand{\Ab}{\mbox{\boldmath $A$}}
\newcommand{\Abdag}{\mbox{\boldmath $A$}^{\dagger}}
\newcommand{\Bb}{\mbox{\boldmath $B$}}

\newcommand{\del}{\partial}

\newcommand{\beq}{\begin{equation}}
\newcommand{\beqa}{\begin{eqnarray}}
\newcommand{\eeq}{\end{equation}}
\newcommand{\eeqa}{\end{eqnarray}}

\newcommand{\Mc}{{\cal M}}

\newcommand{\Udag}{U^{\dagger}}
\newcommand{\Vdag}{V^{\dagger}}

\title{
Microscopic Derivation of Collective Hamiltonian by Means of
the Adiabatic Self-Consistent Collective Coordinate Method \\
{\it ----- Shape Mixing in Low-lying States of ~$^{68}$Se and $^{72}$Kr -----}
}

\author{Nobuo HINOHARA,$^1$~~ Takashi NAKATSUKASA,$^2$~~ 
Masayuki MATSUO$^3$ \\ 
and~ Kenichi MATSUYANAGI$^1$}
\inst{
{\small\it $^1$ Department of Physics, Graduate School of Science,}\\
{\small\it Kyoto University, Kyoto 606-8502, Japan}\\
{\small\it $^2$ Theoretical Nuclear Physics Laboratory, RIKEN Nishina Center,}\\
{\small\it Wako 351-0198, Japan} \\
{\small\it $^3$ Department of Physics, Faculty of Science,}\\
{\small\it Niigata University, Niigata 950-2181, Japan}
}
\abst{
Microscopic dynamics of the oblate--prolate shape coexistence/mixing  
phenomena in $^{68}$Se and $^{72}$Kr are studied by means of 
the adiabatic self-consistent collective coordinate (ASCC) method 
in conjunction with the pairing--plus--quadrupole (P+Q) Hamiltonian 
including the quadrupole pairing interaction. 
Quantum collective Hamiltonian is constructed, and 
excitation spectra, spectroscopic quadrupole moments and 
quadrupole transition properties are evaluated. 
The effect of the time--odd pair field on the collective mass 
(inertia function) of the large-amplitude vibration and the rotational 
moments of inertia about three principal axes is evaluated. 
Basic properties of the shape coexistence/mixing are well reproduced.
The calculation indicates that the oblate--prolate shape mixing decreases 
as the angular momentum increases.  }
\begin{document}

\maketitle

\section{Introduction}
Microscopic understanding of nuclear collective dynamics 
is one of the goals of nuclear structure theory. 
The quasiparticle random phase approximation (QRPA) based on the 
Hartree--Fock--Bogoliubov (HFB) mean-field is the well known 
theoretical approach to the collective dynamics,
but it is applicable only to small--amplitude collective motions 
around the local minima of the potential energy surface.
\cite{rin80,bla86,abe83,bri05,BMtextII}
Nuclei exhibit a variety of large-amplitude collective process
such as anharmonic vibrations, shape coexistence, and fission.
Though the challenge to construct microscopic theory of 
large-amplitude collective motion has long history,
\cite{row76,bri76,vil77,mar77,bar78,goe78,mar80,gia80,dob81,goe81,muk81,
row82,fio83,rei84,kur83,yam84,mat85,mat86,shi87,yam87,
bul89,wal91,kle91a,kan94,nak98a,nak98b,nak00,lib99,yul99,pro04,alm04a,alm04b,
alm05a,alm05b,kle91b,dan00,kur01} some serious problems remain unsolved. 

The self-consistent collective coordinate (SCC) method~\cite{mar80,mat86}
is a microscopic theory of large-amplitude collective motion.
This method with the $(\eta^\ast, \eta)$ expansion technique enables 
us to extract the collective variables from 
the huge-dimensional phase space associated with 
the time-dependent Hartree--Fock--Bogoliubov (TDHFB) state vectors,
and to derive the collective Hamiltonian 
starting from a microscopic Hamiltonian.
The SCC method has been successfully applied to various kinds
of non-linear phenomena in nuclei; such as
anharmonic $\gamma$-vibrations,\cite{mat84,mat85a,mat85b,mat92}
shape phase transitions, \cite{tak89,yam89a,yam89b,yam91}
two-phonon states \cite{aib90} and 
collective rotations.\cite{ter91, ter92,shi01}.
The $(\eta^\ast, \eta)$ expansion is not suitable, however,  
for large-amplitude collective motions 
like shape coexistence/mixing phenomena,
where microscopic description of many-body tunneling effect 
between different local minima in the collective potential energy 
surface becomes the major task. 

The adiabatic SCC (ASCC) method \cite{mat00} is an alternative way to solve
the basic equations of the SCC method assuming that the large-amplitude 
collective motion of interest is slow (adiabatic). Under this assumption, 
the basic equations of the SCC method are expanded up to second order 
with respect to the collective momentum, but the collective coordinate is 
treated non-perturbatively.
Quite recently, we have given a rigorous formulation of the ASCC 
method in which the gauge invariance with respect to the particle number 
fluctuation degrees of freedom is taken into account.\cite{hin07} 

As is well known, in contrast to the remarkable progress in the calculation of 
the collective potential energy, the present status of the microscopic theory 
is quite unsatisfactory concerning the evaluation of the collective mass 
(inertia function) associated with the collective kinetic energy. 
Although the Inglis--Belyaev cranking mass is widely used, 
it violates the self-consistency by ignoring the effect of 
the time--odd component of the moving mean-field.\cite{rin80}
The time--odd mean-field effect is taken into account in the 
collective mass derived by the QRPA, but its application 
is restricted to small--amplitude collective motions around the equilibrium 
states. Concerning large-amplitude collective motions, 
though the effect of the time--odd component generated by the residual 
particle--hole interaction was investigated in a few decades ago,\cite{dob81}
the time--odd effect generated by the residual pairing interaction
has not been discussed so far.
Quite recently, we have shown, using the ASCC method in conjunction with 
the schematic model Hamiltonian,\cite{mat82,miz81,suz88,fuk91} 
that the time-odd pair field increases the collective mass.\cite{hin06}. 
It remains to be seen, however, how it affects the shape coexistence dynamics 
discussed below.

Let us turn to the recent experimental data we are interested in.
Nuclei along the $N=Z$ line evolves its shape drastically 
with changing number of protons and neutrons. 
\cite{woo92,fis00,fis03,bou03,gad05}
The HFB calculation\cite{yam01} indicates that various shapes will 
appear along the $N=Z$ lines: a triaxial ground state for $^{64}$Ge, 
oblate ground states for $^{68}$Se and $^{72}$Kr, 
strongly deformed prolate ground states for $^{76}$Sr, $^{80}$Zr 
and $^{84}$Mo. 
Furthermore, oblate and prolate states may coexist in these nuclei 
except $^{64}$Ge. 
In $^{68}$Se and $^{72}$Kr, the ground and excited states  
corresponding to the oblate and prolate shapes have been found experimentally.
\cite{fis00,fis03,bou03}
From the viewpoint of collective dynamics based on the mean-field theory, 
it is expected that the oblate and prolate shapes are
mixed by the many-body tunneling effect through 
the potential barrier lying between the two local minima in the potential 
energy landscape.
The low-lying states in $^{68}$Se and $^{72}$Kr have been investigated 
by various theoretical approaches beyond the mean-field approximation: 
Large-scale shell model calculation for $^{68}$Se using the pfg-shells 
outside the $^{56}$Ni core,\cite{kan04} 
shell model Monte Carlo calculation for $^{72}$Kr employing the 
pf-sdg shells,\cite{lan03} 
configuration mixing calculation for $^{72-78}$Kr on the basis of 
the particle number and angular momentum projected  
generator coordinate method,\cite{ben06}
and Excited Vampir variational calculation for $^{68}$Se and $^{72}$Kr. 
\cite{pet02,pet96,pet00}. 
Quite recently, Almehed and Walet\cite{alm04a,alm04b,alm05a,alm05b} 
discussed collective paths connecting the oblate and prolate minima 
in $^{68}$Se and $^{72-78}$Kr 
by means of the approach similar to the ASCC method.

The ASCC method was first tested \cite{kob03} in the schematic model 
\cite{mat82,miz81,suz88,fuk91} and then applied\cite{kob04}
to the oblate--prolate shape coexistence phenomena in $^{68}$Se and
$^{72}$Kr with the use of the pairing--plus--quadrupole (P+Q) Hamiltonian.
\cite{bes69,bar65,bar68,kum67}
In both nuclei, the one-dimensional collective path connecting the two
potential local minima is extracted. It was found that the collective path 
runs approximately along the valley of the potential energy surface lying in
the triaxial deformed region. This indicates that the triaxial degree
of freedom is essential for the description of large-amplitude shape
mixing in $^{68}$Se and $^{72}$Kr. 
In the previous work,\cite{kob04} however, 
requantization of the collective Hamiltonian was not done,  
and excitation spectra, electromagnetic transition probabilities, and 
shape mixing probabilities in individual eigen-states were not evaluated. 

This paper presents the result of the first application of the gauge 
invariant formulation\cite{hin07} of the ASCC method to nuclear structure 
phenomena. Thus, the major effort is directed to examine the feasibility of 
the gauge-invariant ASCC method for describing the shape coexistence/mixing 
phenomena. Hereafter we call this new version ``ASCC method'' 
dropping the adjective `` gauge-invariant'' for simplicity.  
More detailed investigation of experimental data and comparison with 
other approaches are planned for future.  
We derive the quantum collective Hamiltonian that describes the coupled 
collective motion of the large-amplitude vibration responsible for the 
oblate--prolate shape mixing and the three-dimensional rotation of the 
triaxial shape. To evaluate the rotational moments of inertia, 
we extend the well-known QRPA equation for rotational motion, 
which yields the Thouless--Valatin moment of inertia, \cite{tho62}
to the non-equibrium states that are defined in the moving-frame 
associated with the large-amplitude vibrational motion.
To clarify the role of the time-odd pair field in shape
mixing dynamics, we investigate, with the use of the P+Q Hamiltonian 
including the quadrupole pairing interaction, 
its effects on the collective mass of large-amplitude vibration, 
the rotational moments of inertia, 
the energy spectra, transition probabilities, 
and shape mixing probabilities in individual eigen-states.

This paper is organized as follows: 
The basic equations of the ASCC method are summarized in \S\ref{sec:ascc}.
Quasiparticle representation of the microscopic Hamiltonian is 
given in \S\ref{sec:model}. 
Solving procedure of the ASCC equations is presented in 
\S\ref{sec:solution}. 
Collective Schr\"odinger equation is derived in 
\S\ref{sec:quantum}. 
Results of numerical calculation for energy spectra, 
spectroscopic quadrupole moments and quadrupole transition probabilities 
of low-lying states in $^{68}$Se and $^{72}$Kr are presented and discussed 
in \S\ref{sec:result}.
Concluding remarks are given in \S\ref{sec:conclusion}.

A preliminary version of this work was previously reported 
in Ref.~\citen{hin07-2}.

\section{The ASCC method} \label{sec:ascc}

\subsection{Basic equations of the ASCC method}

We first recapitulate the basic equations of the ASCC method.
The TDHFB state $\ket{\phi(t)}$ is written in terms of the collective
variables as 
\begin{align}
 \ket{\phi(t)} = \ket{\phi(q,p,\bm{\varphi},\bm{n})} =
 e^{-i\sum_{\tau}\varphi^{(\tau)} \Nt^{(\tau)}} \ket{\phi(q,p,\bm{n})},
\end{align}
where $q$ and $p$ represent the one-dimensional collective
coordinate and collective momentum, respectively. The variables 
$\bm{\varphi}=(\varphi^{(n)},\varphi^{(p)})$
and 
$\bm{n} = (n^{(n)}, n^{(p)})$
denote the gauge angles in particle number space and number
fluctuations, respectively, which correspond to the canonical coordinates 
and momenta of the pairing rotation restoring the particle number
conservation broken by the HFB approximation.
The operator $\Nt^{(\tau)}\equiv\Nhat^{(\tau)} - N^{(\tau)}_0$ represents 
the particle number measured from the reference value
$N^{(\tau)}_0$, which is set to the number of the valence protons ($\tau=p$) 
and neutrons ($\tau=n$) in the model space.

The intrinsic state with respect to the pairing rotation is written as
$\ket{\phi(q,p,\bm{n})}=e^{i\Ghat(q,p,\bm{n})}\ket{\phi(q)}$, where 
$\ket{\phi(q)}\equiv\ket{\phi(q,p=0,\bm{n}={\textbf 0})}$.
Assuming that large-amplitude collective motion is adiabatic, 
that is, the collective momentum $p$ and the number fluctuation $\bm{n}$
are small, we expand the one-body operator $\Ghat(q,p,\bm{n})$ 
with respect to $p$ and $n^{(\tau)}$ and consider only the first order: 
\begin{align}
 \ket{\phi(q,p,\bm{n})} = e^{ip\Qhat(q) +
 i\sum_{\tau}n^{(\tau)}\That^{(\tau)}(q)} \ket{\phi(q)}, \label{eq:TDHFBvector}
\end{align}
where $\Qhat(q)$ is a time-even one-body operator, while
$\That^{(\tau)}(q)$ is a time-odd one-body operator. 
Using quasiparticle creation and annihilation operators, 
$\adag_{\alpha}(q)$ and $a_{\alpha}(q)$, defined with respect to 
a moving-frame HFB state $\ket{\phi(q)}$, which satisfy the condition 
$a_{\alpha}(q) \ket{\phi(q)} = 0$, they are written as 
\begin{align}
 \Qhat(q) &= \Qhat^A(q) + \Qhat^B(q) \nonumber \\
          &= \sum_{\alpha\beta}\left( Q^A_{\alpha\beta}(q)
 \adag_{\alpha}(q) \adag_{\beta}(q) + Q^{A\ast}_{\alpha\beta}(q)
 a_{\beta}(q) a_{\alpha}(q) + Q^B_{\alpha\beta}(q)
 \adag_{\alpha}(q) a_{\beta}(q)\right), \\
 \That^{(\tau)}(q) &= \sum_{\alpha\beta}\left( \Theta^{(\tau)A}_{\alpha\beta}(q)
 \adag_{\alpha}(q) \adag_{\beta}(q) +\Theta^{(\tau)B\ast}_{\alpha\beta}(q) a_{\beta}(q) a_{\alpha}(q)\right).
\end{align}
Note that the operator $\Qhat(q)$ contains the $B$-part 
(third term in the r.h.s.), in addition to the $A$-part
(first and second terms) in order to satisfy the gauge-invariance of 
the ASCC equations.~\cite{hin07}
In the following we will omit the index $q$ in the quasiparticle
operators for simplicity of notation.

The basic equations of the SCC method consists of the canonical variable
conditions, the moving-frame HFB equation, and the moving-frame QRPA equations.
Below we summarize the lowest order expressions of these equations 
with respect to the expansion in  $p$ and $\bm{n}$ (See Ref.~\citen{hin07} 
for their derivations).
The canonical variable conditions are given by 
\begin{align}
 \bra{\phi(q)}\Phat(q)\ket{\phi(q)} &= 0, \label{eq:can-p} \\
 \bra{\phi(q)}\Qhat(q)\ket{\phi(q)} &= 0,\label{eq:can-q} \\
 \bra{\phi(q)}\Nt^{(\tau)}\ket{\phi(q)}    &= 0,\label{eq:can-n} \\
 \bra{\phi(q)}\That^{(\tau)}(q)\ket{\phi(q)} &= 0,\label{eq:can-t} \\
 \bra{\phi(q)}[\That^{(\tau)}(q),\Nt^{(\tau')}]\ket{\phi(q)} &= i\delta_{\tau\tau'}, \label{eq:can-nt}\\
 \bra{\phi(q)}[\Qhat(q),\That^{(\tau)}(q)]\ket{\phi(q)} &= 0, \label{eq:can-qt}\\ 
 \bra{\phi(q)}\frac{\del \Qhat}{\del q}\ket{\phi(q)} &= -1,  \label{eq:can-qdel}
\end{align}
where $\Phat(q)$ is the local shift operator defined by
\begin{align}
 \Phat(q)\ket{\phi(q)} = i\frac{\del}{\del q}\ket{\phi(q)}. \label{eq:defP}
\end{align}
Differentiating (\ref{eq:can-q}) and (\ref{eq:can-n}) with respect to 
$q$ and using (\ref{eq:can-qdel}) and (\ref{eq:defP}), we obtain 
\begin{align}
 \bra{\phi(q)}[\Qhat(q),\Phat(q)]\ket{\phi(q)} &= i, \label{eq:can-qp} \\
 \bra{\phi(q)}[\Nt^{(\tau)}, \Phat(q)]\ket{\phi(q)} &= 0.  \label{eq:can-np}
\end{align}
Equations (\ref{eq:can-p}), (\ref{eq:can-q}), (\ref{eq:can-n}) and
(\ref{eq:can-t}) assure that the constant term of those operators 
are zero in their quasiparticle representations, while
Eqs.~(\ref{eq:can-nt}), (\ref{eq:can-qt}), (\ref{eq:can-qp}) and 
(\ref{eq:can-np}) guarantee orthonormalization of 
the collective mode and the number fluctuation modes.
Equation (\ref{eq:can-qdel}) defines the scaling of the collective coordinate.

The moving-frame HFB equation is given by 
\begin{align}
 \delta\bra{\phi(q)}\Hhat_M(q)\ket{\phi(q)} = 0, \label{eq:ascc1}
\end{align}
where
\begin{align}
 \Hhat_M(q) = \Hhat - \sum_{\tau}\lambda^{(\tau)}(q)\Nt^{(\tau)} - \frac{\del V}{\del q}\Qhat(q)
 \label{eq:ascc1b}
 \end{align}
represents the moving-frame Hamiltonian 
with the chemical potential $\lambda^{(\tau)}(q)$ 
and the collective potential $V(q)$ defined by 
 \begin{align} 
 \lambda^{(\tau)}(q) =& \frac{\del \Hc}{\del n^{(\tau)}}\Big\arrowvert_{p=0,\bm{n}={\textbf 0},\vec{I}=\vec{0}} =
 \bra{\phi(q)}[\Hhat, i\That^{(\tau)}(q)]\ket{\phi(q)}, \label{eq:deflambda}\\
 V(q) =&  \Hc(q,p,\bm{n},\vec{I})\Big\arrowvert_{p=0,\bm{n}={\textbf 0},\vec{I}=\vec{0}} =
 \bra{\phi(q)}\Hhat\ket{\phi(q)}. 
\end{align}
 
The moving-frame QRPA equations are given by 
\begin{align}
\delta\bra{\phi(q)}[\Hhat_M(q), i\Qhat(q) ] - B(q)\Phat(q)
\ket{\phi(q)} = 0, \label{eq:ascc2}
\end{align}
\begin{multline}
\delta\bra{\phi(q)} [\Hhat_M(q), \Phat(q)] - iC(q)\Qhat(q) \\
- \frac{1}{2B(q)}[[\Hhat_M(q), \frac{\del V}{\del q}\Qhat(q)],
 i\Qhat(q)]
- i\sum_{\tau} \frac{\del\lambda^{(\tau)}}{\del q} \Nt^{(\tau)} \ket{\phi(q)} = 0,
\label{eq:ascc3}
\end{multline}
where $B(q)$ and $C(q)$ represent the inverse collective mass and 
the local stiffness, respectively. They are defined by 
\begin{align}
 B(q) =& \frac{\del^2 \Hc}{\del p^2}
 \Big\arrowvert_{p=0,\bm{n}={\textbf 0},\vec{I}=\vec{0}} =
 \bra{\phi(q)}[[\Hhat, i\Qhat(q)], i\Qhat(q)]\ket{\phi(q)}, \label{eq:defB}
\end{align}
\begin{align}
 C(q) = \frac{\del^2 V}{\del q^2}
 + \frac{1}{2B(q)}\frac{\del B}{\del q}\frac{\del V}{\del q}.
\end{align}

Note that the ASCC equations, (\ref{eq:ascc1}), (\ref{eq:ascc2}) and
(\ref{eq:ascc3}), are invariant against the following transformation~\cite{hin07}
\begin{align}
 \Qhat(q) \rightarrow &\, \Qhat(q) + \alpha^{(\tau)} \Nt^{(\tau)}, \nonumber \\
 \lambda^{(\tau)}(q) \rightarrow &\, \lambda^{(\tau)}(q) - \alpha^{(\tau)} \frac{\del
 V}{\del q}(q), \nonumber \\
 \frac{\del\lambda^{(\tau)}}{\del q}(q) \rightarrow &\,
 \frac{\del\lambda^{(\tau)}}{\del q}(q) - \alpha^{(\tau)} C(q). \label{eq:transform}
\end{align}
Therefore it is necessary to fix the particle number gauge for neutrons
and protons in order to find the unique solution of the ASCC equations.
The algorithm to find simultaneous solutions of Eqs.~(\ref{eq:ascc1}), 
(\ref{eq:ascc2}) and (\ref{eq:ascc3}) fulfilling 
the canonical variables conditions and 
the gauge-fixing condition is described in \S\ref{sec:solution}.

In this paper, we take into account the rotational motion 
as well as the large-amplitude vibrational motion by 
considering the collective Hamiltonian defined as follows: 
\begin{align}
 \Hc(q,p,\bm{n},\vec{I}) =& \bra{\phi(q,p,\bm{n})} \Hhat
 \ket{\phi(q,p,\bm{n})} + \sum_{i=1}^{3}\frac{1}{2\Jc_i(q)}I_i^2
 \nonumber \\
=& V(q) + \frac{1}{2}B(q)p^2 + \sum_{\tau} \lambda^{(\tau)}(q)n^{(\tau)}
+ \sum_{i=1}^3 \frac{1}{2\Jc_i(q)}I_i^2. \label{eq:Hcoll}
\end{align}
The first and the second terms represent the potential and kinetic energies 
of the large-amplitude collective vibration, respectively, while  
the third and the fourth terms the energies associated with  
the particle--number fluctuations and 
the three-dimensional rotations of triaxially deformed mean fields, 
respectively. 
The three rotational moments of inertia $\Jc_i(q)$ are defined 
with respect to the principal axes 
associated with the moving-frame HFB state $\ket{\phi(q)}$ and evaluated by 
\begin{gather}
\delta \bra{\phi(q)} [\Hhat_M(q), \Psihat_i(q)]
 - \frac{1}{i}\Jc^{-1}_i(q)\Ihat_i 
\ket{\phi(q)} = 0, \label{eq:ascc-rot} \\
\bra{\phi(q)}[\Psihat_i(q), \Ihat_j]\ket{\phi(q)} = \delta_{ij}, 
\label{eq:can-qI}
\end{gather}
where $\Psihat_i(q)$ and $\Ihat_i$ represent 
the rotational angle and the angular momentum operators, respectively.
These equations reduce to the well-known QRPA equations giving 
the Thouless--Valatin moments of inertia~\cite{tho62} 
when $\ket{\phi(q)}$ is an equilibrium state 
corresponding to a local minimum of the collective potential energy $V(q)$.
We call them ``Thouless--Valatin equations,''
although they are in fact extensions of the QRPA equations 
for collective rotation to non-equilibrium HFB states $\ket{\phi(q)}$.
Note that $\Hhat_M(q)$ appears in Eq.~(\ref{eq:ascc-rot}) 
instead of $\Hhat$.

\section{Hamiltonian} \label{sec:model}

We adopt the Hamiltonian consisting of the spherical single-particle energy, 
the monopole and the quadrupole pairing interactions, 
and the quadrupole particle-hole interaction:
\begin{align}
\Hhat =&  \sum_k \varepsilon_k \cdag_k c_k
- \sum_{\tau} \frac{G_0^{(\tau)}}{2} (\Atdag \Ahat^{(\tau)} + \Ahat^{(\tau)} \Atdag)
\nonumber \\
&- \sum_{\tau} \frac{G_2^{(\tau)}}{2}
\sum_{K=-2}^{2} 
( \Btdag_{2K} \Bhat^{(\tau)}_{2K} + \Bhat^{(\tau)}_{2K} \Btdag_{2K})
- \frac{\chi}{2} \sum_{K=-2}^{2} \Ddag_{2K} \Dhat_{2K},
\label{eq.hamil}
\end{align}
where the monopole pairing operator $\Atdag$, the quadrupole pairing operator 
$\Btdag_{2K}$, the quadrupole particle-hole operator $\Dhat_{2K}$ are defined by
\begin{align}
 \Atdag &= \sum_{(k,\tilde{k})\in\tau} \cdag_k \cdag_{\tilde{k}}, \\
 \Btdag_{2K} &= \sum_{kl\in\tau} D_{2K}^{(\tau)}(kl) \cdag_k
 \cdag_{\tilde{l}}, \\
 \Dhat_{2K} &= \sum_{\tau=n,p} \sum_{kl\in \tau}
 D_{2K}^{(\tau)}(kl)\cdag_k c_l. \label{eq:quadrupole}
\end{align}
Here $\cdag_k$ is the nucleon creation operator, and $k$ 
denotes the set of the quantum numbers of single-particle
state $(N_k, j_k, l_k, m_k)$.
The operator $\cdag_{\tilde{k}}$ represents its time-reversal state,
\begin{align}
 \cdag_{\tilde{k}} = (-1)^{j_k+m_k} \cdag_{-k},
\end{align}
where the index $-k$ denotes $(N_k, j_k, l_k, -m_k)$.
The quadrupole matrix elements are given by
\begin{align}
 D_{2K}^{(\tau)}(kl) = \alpha_{\tau}^2 \bra{k} r^{2} Y_{2K} \ket{l}, 
\quad (kl \in \tau),
\end{align}
where the factors, $\alpha_n^2= (2N/A)^{2/3}$ and $\alpha_p^2=(2Z/A)^{2/3}$, 
% Check !! different definition in Kobayasi et al and Baranger Kumar II
are multiplied to yield the same root mean square radius for neutrons and 
protons. For $N$=$Z$ nuclei like $^{68}$Se and $^{72}$Kr, 
these factors are unity.
Following Baranger and Kumar \cite{bar68},
we employ a model space consisting of two major oscillator shells 
(the total quantum number of the lower shell is denoted  $N_L$ and 
that of the upper shell $N_L+1$),
and multiply the reduction factor
$\zeta=(N_L+3/2) / (N_L + 5/2)$
to the quadrupole matrix elements $D_{2K}^{(\tau)}(kl)$ of the upper shell.
Following the conventional prescription of the P+Q model, we ignore 
the Fock terms. Accordingly, we use an abbreviation HB in place of HFB 
in the following.

We rewrite Hamiltonian (\ref{eq.hamil}) into the following form
\begin{align}
 \Hhat = \sum_{k} \varepsilon_k \cdag_k c_k
 - \frac{1}{2} \sum_{s}\kappa_s \Fhatp_s \Fhatp_s 
 + \frac{1}{2} \sum_{s}\kappa_s \Fhatm_s \Fhatm_s,
\label{eq:separableH}
\end{align}
where the Hermite operators $\Fhatp_s$ and the anti-Hermite operators $\Fhatm_s$
are defined by
\begin{gather}
\begin{align}
 \Fhat^{(\pm)}_s = \frac{1}{2}& (\Fhat_s \pm \Fhat_s^\dagger), \\
 \Fhat_{s=1-15} = \{ &\Ahat^{(n)}, \Ahat^{(p)},
 \Bhat^{(n)}_{20(+)},
\Bhat^{(n)}_{21(+)}, \Bhat^{(n)}_{21(-)},
\Bhat^{(n)}_{22(+)}, \Bhat^{(n)}_{22(-)}, \nonumber \\
&\Bhat^{(p)}_{20(+)},
\Bhat^{(p)}_{21(+)}, \Bhat^{(p)}_{21(-)},
\Bhat^{(p)}_{22(+)}, \Bhat^{(p)}_{22(-)},
\Dhat_{20} ,\Dhat_{21}, \Dhat_{22}\}.
\end{align}
\end{gather}
Here we use
\begin{align}
 \Btdag_{2K(\pm)} \equiv \frac{1}{2}
(\Btdag_{2K} \pm \Btdag_{2-K}),  \quad (K \ge 0),
\end{align}
in place of $\Btdag_{2K}$ for the quadrupole pairing operators. 
The interaction strengths $\kappa_s$ are given by
\begin{align}
 \kappa_{s=1-15} =  \{ &
 2G_0^{(n)}, 2G_0^{(p)},
2G_2^{(n)}, 4G_2^{(n)}, 4G_2^{(n)}, 4G_2^{(n)}, 4G_2^{(n)}, \nonumber \\
& 2G_2^{(p)}, 4G_2^{(p)}, 4G_2^{(p)}, 4G_2^{(p)}, 4G_2^{(p)},
\chi, 2\chi, 2\chi\}.
\end{align}
This Hamiltonian is invariant against a rotation 
by $\pi$ about the $x$-axis. The quantum number associated with this
is called the signature, $r=e^{-i\pi\alpha}$.
The single-particle basis with definite signatures are defined by
\begin{align}
d_k \equiv \frac{1}{\sqrt{2}} ( c_k + c_{\tilde{k}}), &\quad r = -i \quad
 (\alpha = \frac{1}{2}), \nonumber \\
d_{\bar{k}} \equiv \frac{1}{\sqrt{2}} ( c_{\tilde{k}} - c_k), &\quad
r = i \quad (\alpha = -\frac{1}{2}).
\end{align}
where $k$ denotes the single-particle basis whose magnetic 
quantum number satisfies the condition $m_k-1/2={\rm even}$.
The operators $\Fhat_s^{(\pm)}$ can be classified according to their 
signature and $K$-quantum numbers as shown in Table~\ref{table:quantum-number}.

The large-amplitude collective vibration responsible for the 
oblate--prolate shape mixing is associated with 
the $K=0$ and $2$ components of the interactions in the positive-signature 
($r=+1$) sector.
Thus the infinitesimal generator of large-amplitude collective motion, 
$\Qhat(q)$, can be written in terms of the single-particle basis with 
definite signature as
\begin{align}
 \Qhat(q) = \sum_{\tau} {\sum_{kl\in\tau}}' Q^{(\tau)}_{kl}(q)\ddg_k d_{l} 
 + Q^{(\tau)}_{\bar{k}\bar{l}}(q) \ddg_{\bar{k}} d_{\bar{l}},
\end{align}
where ${\sum}'$ denotes a sum over the signature pairs ($k, \bar{k}$), and
$Q^{(\tau)}_{kl}=Q^{(\tau)}_{\bar{k}\bar{l}}$.
On the other hand, the $K=1$ component of the interaction in the $r=+1$ sector 
and the $K=1$ and $2$ components in the $r=-1$ sector
contribute to Thouless--Valatin equations (\ref{eq:ascc-rot}). 

\begin{table}[htbp]
\begin{center}
\caption{
Classification of one-body operators $\Fhat_s^{(\pm)}$ 
in terms of the signature $r$(or $\alpha$) and $K$ quantum numbers.}
\label{table:quantum-number}
\begin{tabular}{ccc} \hline \hline
 & $r = +1 (\alpha=0)$ & $r = -1 (\alpha=1)$ \\ \hline
$K=0$ &
 $\{\Ahat_n^{(\pm)},\Ahat_p^{(\pm)},\Bhat_{20(+)}^{(n)(\pm)},\Bhat_{20(+)}^{(p)(\pm)},\Dhat_{20}^{(+)}\}$ & $-$
 \\
$K=1$ &
$\{\Bhat_{21(-)}^{(n)(\pm)}, \Bhat_{21(-)}^{(p)(\pm)},
 \Dhat_{21}^{(-)}\}$ &
$\{\Bhat_{21(+)}^{(n)(\pm)}, \Bhat_{21(+)}^{(p)(\pm)},
 \Dhat_{21}^{(+)}\}$  \\
$K=2$ &
$\{\Bhat_{22(+)}^{(n)(\pm)}, \Bhat_{22(+)}^{(p)(\pm)},
 \Dhat_{22}^{(+)}\}$ &
$\{\Bhat_{22(-)}^{(n)(\pm)}, \Bhat_{22(-)}^{(p)(\pm)},
 \Dhat_{22}^{(-)}\}$
 \\ \hline
\end{tabular}
\end{center}
\end{table}

\section{Solution of the ASCC equations for separable interactions} 
\label{sec:solution}

\subsection{The ASCC equations for separable interactions}

For the separable interactions (\ref{eq:separableH}), the ASCC equations are 
written as \cite{mat00,hin07}
\begin{align}
\delta\bra{\phi(q)}\hhat_M(q) \ket{\phi(q)} = 0,
\label{eq:ascc1_sep}
\end{align}
\begin{align}
\delta\bra{\phi(q)}[\hhat_M(q), \Qhat(q) ] - \sum_s \fm_{Q,s}(q)
 \Fhatm_s
- \frac{1}{i} B(q) \Phat(q)
     \ket{\phi(q)} = 0,
\label{eq:ascc2_sep}
\end{align}
\begin{align}
\delta\bra{\phi(q)} & \left[\hhat_M(q), \frac{1}{i}B(q)\Phat(q)\right]
    - \sum_s \fp_{P,s}(q) \Fhatp_s
    - \omega^2(q) \Qhat(q) \nonumber \\
    &- \sum_s f^{(+)}_{R,s}(q) \Fhatp_s
-\frac{1}{2}[[\hhat_M(q), \frac{\del V}{\del q}\Qhat(q)], \Qhat(q)]
 \nonumber \\
    &+ \sum_s \left[\Fhatm_s, \frac{\del V}{\del q}\Qhat(q)\right]
 f^{(-)}_{Q,s}(q)
    - \sum_{\tau} f_{N}^{(\tau)}(q) \Nt^{(\tau)}
    \ket{\phi(q)} =0,
\label{eq:ascc3_sep}
\end{align}
where $\omega^2(q)=B(q)C(q)$ is the moving-frame QRPA frequency squared, and
$\hhat_M(q)$ denotes the self-consistent mean-field Hamiltonian
in the moving frame, defined by
\begin{align}
\hhat_M(q) = \hhat(q)  - \sum_{\tau} \lambda^{(\tau)}(q)\Nt^{(\tau)}
                       - \frac{\del V}{\del q}\Qhat(q),
\end{align}
with
\begin{align}
\hhat(q) &= \hhat_0
- \sum_s \kappa_s \Fhatp_s \bra{\phi(q)}\Fhatp_s\ket{\phi(q)} .
\end{align}
In the above equations, the summation over $s$
is restricted to the operators with $K=0$ and $2$ 
in the positive-signature sector.
We also define the following quantities
\begin{subequations}
\begin{align}
 &\fm_{Q,s}(q) = -\kappa_s
              \bra{\phi(q)}[\Fhatm_s, \Qhat(q)] \ket{\phi(q)},
 \label{eq:fQs}\\
 &\fp_{P,s}(q)=
   \kappa_s \bra{\phi(q)}[\Fhatp_s,\frac{1}{i} B(q)\Phat(q)]
 \ket{\phi(q)},
   \label{eq:fPs}\\
 &\fp_{R,s}(q)= -\frac{1}{2}\kappa_s
 \bra{\phi(q)}\left[\left[\Fhatp_s,\frac{\del V}{\del
 q}\Qhat(q)\right],\Qhat(q)\right]
 \ket{\phi(q)}, \label{eq:fRs}\\
 &f_{N}^{(\tau)}(q) = B(q)\frac{\del \lambda^{(\tau)}}{\del q}. \label{eq:fN}
\end{align}
\label{eq:f-mfQRPA}
\end{subequations}
Note that all matrix elements are real and 
$\bra{\phi(q)}\Fhatm_s\ket{\phi(q)}=0$.

\subsection{Overview of solving procedure}
\label{subsec:algorithm}
The infinitesimal generators, $\Qhat(q)$ and $\Phat(q)$,
which are represented on top of the quasiparticle vacuum $\ket{\phi(q)}$,
are the solutions of the moving-frame QRPA equations,
while the quasiparticle vacuum $\ket{\phi(q)}$, which depends
on $\Qhat(q)$, is a solution of the moving-frame HB equation.
In order to construct the collective path, we have to 
obtain a self-consistent solution for the quasiparticle vacuum 
and the infinitesimal generators. This requires a double 
iterative procedure for each value of $q$, because 
the moving-frame HB equation is also solved by iteration.
\begin{description}

\item{{\it Step 0: Starting point}}

The shape coexistence phenomena imply that several solutions 
of the static HB equation exist representing different local minima
in the potential energy surface. 
We can choose one of the HB solutions and assume that it is on the collective 
path.  This starting state is denoted as $\ket{\phi(q=0)}$.
In the calculation for $^{68}$Se and $^{72}$Kr in this paper, 
we choose the HB state at the lowest minimum, which possess the oblate shape. 
As discussed in Ref.~\citen{hin07}, a gauge fixing is necessary to solve 
the moving-frame QRPA equations. We choose the ``ETOP'' gauge.

\item{{\it Step 1: Initial setting}}

Assume that the solution of the ASCC equations at $q-\delta q$ is obtained.
In order to calculate the solution at $q$, 
we start from solving the moving-frame HB equation (\ref{eq:ascc1_sep}).
As an infinitesimal generator in the moving-frame Hamiltonian, 
we use an initial trial generator $\Qhat(q)^{(0)}$ 
constructed from the 
lowest two solutions of the moving-frame QRPA equations at $q-\delta q$ of the 
following form
\begin{align}
 \Qhat(q)^{(0)} = (1-\varepsilon) \Qhat_1(q -\delta q) + \varepsilon
 \Qhat_2(q-\delta q),
\end{align}
where $\Qhat_1(q-\delta q)$ and $\Qhat_2(q-\delta q)$ 
denote the lowest and the second--lowest solutions of the moving-frame 
QRPA equations at $q-\delta q$, respectively. 
The mixing parameter $\varepsilon$ is set to 0.1.
This choice is crucial 
to find a symmetry breaking solution in the moving-frame QRPA equations 
when the moving-frame HB state $\ket{\phi(q-\delta q)}$ and the moving-frame QRPA mode
$\Qhat_1(q-\delta q)$ hold the axial symmetry.\cite{kob04}
The $\delta q$ is set to 0.0157 in the present calculation.

\item{{\it Step 2: Solving the moving-frame HB equation}}

Using the operator $\Qhat^{(n-1)}(q) \, (n \ge 1)$, 
solve the moving-frame HB equation at $q$
\begin{align}
 \delta \bra{\phi^{(n)}(q)}
\Hhat - \sum_{\tau}(\lambda^{(\tau)}(q))^{(n)}\Nt^{(\tau)} - \frac{\del V}{\del q}^{(n)}(q)
 \Qhat^{(n-1)}(q)
\ket{\phi^{(n)}(q)} = 0,
\end{align}
with three constraints from the canonical variable conditions
\begin{align}
 \bra{\phi^{(n)}(q)}\Nt^{(\tau)}\ket{\phi^{(n)}(q)} &= 0, \\
 \bra{\phi^{(n)}(q)}\Qhat(q-\delta q)\ket{\phi^{(n)}(q)}& = \delta q.
\end{align}
This step is discussed in \S\ref{sec:mfhb} in detail.

\item{{\it Step 3: Solving the moving-frame QRPA equations}}

Using the moving-frame HB state $\ket{\phi^{(n)}(q)}$,
the Lagrange multipliers $(\lambda^{(\tau)}(q))^{(n)}$ and 
$\del V/\del q (q)^{(n)}$
obtained in the
previous step, we solve the moving-frame QRPA equations
with the same gauge-fixing condition as used at the HB state in {\it Step 0}.
This determines the infinitesimal generator $\Qhat^{(n)}(q)$ 
as the lowest solution of Eqs.~(\ref{eq:ascc2_sep}) and (\ref{eq:ascc3_sep}).
Details of this step are described in \S\ref{sec:mfQRPA} and 
Appendix~\ref{app:Bpart}.

\item{{\it Step 4: Achieving the self-consistency}}

Updating the operator $\Qhat^{(n)}(q)$, we go back to {\it Step 2}, and
repeat {\it Steps 2} and {\it 3} until all quantities at $q$ converge.

\item{{\it Step 5: Progression}}

Change $q$ to $q+\delta q$ and return to {\it Step 1}.

\end{description}

Carrying out {\it Steps 1-5}, we obtain a collective path starting from 
the HB minimum to one direction $(q>0)$. We then change the sign of $\delta q$ 
and repeat the above procedure to the opposite direction ($q < 0$). 
In this way we obtain an entire collective path.  

After obtaining the solutions of the ASCC equations, 
we solve the Thouless--Valatin eq.~(\ref{eq:ascc-rot}) 
at every point on the collective path using the moving-frame HB state 
$\ket{\phi(q)}$ 
to evaluate the rotational moments of inertia $\Jc_i(q)$.
Details of this calculation are described in Appendix~\ref{app:MoI}.

\subsection{The moving-frame HB equation in the quasiparticle representation}
\label{sec:mfhb}

The quasiparticle operators $\adag_{\mu}(q)$ and $a_{\mu}(q)$ 
associated with the moving-frame HB state $\ket{\phi(q)}$ are written 
in terms of the nucleon operators, $\ddg_k$ and $d_{\bar{k}}$, 
with definite signature, as
\begin{align}
 \begin{pmatrix}
  \adag_{\mu}(q) \\ a_{\bar{\mu}}(q)
 \end{pmatrix}
 = {\sum_k}'
 \begin{pmatrix}
  U_{\mu k}(q) & V_{\mu\bar{k}}(q) \\
  V_{\bar{\mu}k}(q) & U_{\bar{\mu}\bar{k}}(q)
 \end{pmatrix}
 \begin{pmatrix}
  \ddg_{k} \\ d_{\bar{k}}.
 \end{pmatrix}. \label{eq:qp-trans}
\end{align}
Its inverse transformation is 
\begin{align}
 \begin{pmatrix}
  \ddg_{k} \\ d_{\bar{k}}
 \end{pmatrix}
= {\sum_\mu}'
\begin{pmatrix}
 U_{k\mu}(q) & V_{k\bar{\mu}}(q) \\
 V_{\bar{k}\mu}(q) & U_{\bar{k}\bar{\mu}}(q)
\end{pmatrix}
 \begin{pmatrix}
  \adag_{\mu}(q) \\ a_{\bar{\mu}}(q)
 \end{pmatrix}. \label{eq:qp-trans-inv}
\end{align}
The $U$ and $V$ matrices are determined by solving the moving-frame HB 
equation (\ref{eq:ascc1_sep}). Note that superscripts $\tau (=n,p)$ 
for $U$, $V$, and Fermion operators are omitted for simplicity of notation.

The moving-frame Hamiltonian is written as
\begin{align}
 \hhat_M(q) = \sum_{\tau} {\sum_{kl\in\tau}}'
 \left( (h_{M}^{(\tau)})_{kl}(q)(\ddg_k d_l + \ddg_{\bar{k}} d_{\bar{l}})
 - \Delta_{k\bar{l}}^{(\tau)}(q) (\ddg_k \ddg_{\bar{l}} + d_{\bar{l}} d_k)
 \right),
\end{align}
where the particle-hole part and the particle-particle part of the 
moving-frame Hamiltonian are given by
\begin{align}
  (h_{M}^{(\tau)})_{ll'}(q) =& h^{(\tau)}_{ll'}(q) - \lambda^{(\tau)}(q)
 \delta_{ll'} - \frac{\del V}{\del q}(q) Q^{(\tau)}_{ll'}(q), \\
 h^{(\tau)}_{ll'} =& \varepsilon^{(\tau)}_{l}\delta_{ll'}
 -\sum_{s\in{\rm ph}}
 \kappa_s\bra{\phi(q)}\Fhatp_s\ket{\phi(q)}(l|\Fhatp_s|l'), 
\\
 \Delta^{(\tau)}_{l\bar{l'}} =& \sum_{s\in{\rm pp,hh}} \kappa_s
 \bra{\phi(q)}\Fhatp_s\ket{\phi(q)}(0|\Fhatp_s|l\bar{l'}).
\end{align}
The matrix elements $(k|\Fhatp_s|l)$ are defined by 
\begin{align}
(l|\Fhatp_s|\bar{l'}) = (0|d_l\Fhatp_s \ddg_{\bar{l}}|0), \quad 
(0|\Fhatp_s|l\bar{l'})= (0| \Fhatp_s \ddg_{l} \ddg_{\bar{l'}}|0),
\end{align}
where $|0)$ denotes the vacuum for nucleon operators.

The moving-frame HB equation are thus written
\begin{align}
 {\sum_{ll'\in\tau}}'\left(
 (h^{(\tau)}_{M})_{ll'}(q) U_{l'k}(q) +
 \Delta^{(\tau)}_{ll'}(q)V_{l'k}(q)\right) =& E^{(\tau)}_k U_{lk}(q),
 \\
 {\sum_{ll'\in\tau}}'\left(
 \Delta^{(\tau)}_{ll'}(q) U_{l'k}(q)
 + (h^{(\tau)}_{M})_{ll'}(q) V_{l'k}(q)\right)
 =& -E^{(\tau)}_k V_{lk}(q),
\end{align}
where $E^{(\tau)}_k$ denotes the quasiparticle energy.
These equations are solved under the following three constraints: 
\begin{align}
 \bra{\phi(q)}\Nhat^{(n)}\ket{\phi(q)} =&\ N^{(n)}_{0}, \\
 \bra{\phi(q)}\Nhat^{(p)}\ket{\phi(q)} =&\ N^{(p)}_{0}, \\ 
 \bra{\phi(q)}\Qhat(q-\delta q)\ket{\phi(q)} =&\ \delta q. \label{eq:Q-const}
\end{align}
The Lagrange multipliers, $\lambda^{(n)}(q), \lambda^{(p)}(q)$
and $dV/dq(q)$, are determined such that these constraints are fulfilled. 
The expectation values in the moving-frame Hamiltonian 
are updated using $U_{lk}$ and $V_{lk}$ thus obtained 
until self-consistency is achieved.

In the quasiparticle representation, the moving-frame Hamiltonian 
$\hhat_M(q)$, the neutron and proton number operators $\Nt^{(\tau)}$ 
and the operators $\Fhatpm_s$ with $K=0$ and $2$ in the $r=1$ sector 
are written in the following forms:
\begin{align}
\hhat_M(q) =& \bra{\phi(q)}\hhat_M(q)\ket{\phi(q)}
 + {\sum_{\mu}}'\left( E_{\mu}(q) \Bb_{\mu\mu}(q) + E_{\bar{\mu}}(q)
 \Bb_{\bar{\mu}\bar{\mu}}(q)\right), \\
\Nt^{(\tau)} =& {\sum_{\mu\bar{\nu}}}' N^{(\tau)}_A(\mu\bar{\nu})
 (\Abdag_{\mu\bar{\nu}}(q) + \Ab_{\mu\bar{\nu}}(q))
+ {\sum_{\mu}}' N^{(\tau)}_B(\mu\mu) (\Bb_{\mu\mu}(q) + \Bb_{\bar{\mu}\bar{\mu}}(q)),
\label{eq:N-qp}
\\
\Fhatpm_s =& \bra{\phi(q)}\Fhatpm_s\ket{\phi(q)}
 + {\sum_{\mu\bar{\nu}}}' F^{(\pm)}_{A,s}(\mu\bar{\nu})
(\Abdag_{\mu\bar{\nu}}(q) + \Ab_{\mu\bar{\nu}}(q)) \nonumber \\
 +& {\sum_{\mu\nu}}' F^{(\pm)}_{B,s}(\mu\nu)
(\Bb_{\mu\nu}(q) + \Bb_{\bar{\mu}\bar{\nu}}(q)), \label{eq:F-r=+1,K=0,2}
\end{align}
where
\begin{align}
\Abdag_{\mu\bar{\nu}}(q) = \adag_{\mu}(q) \adag_{\bar{\nu}}(q),\quad
\Ab_{\mu\bar{\nu}}(q) = a_{\bar{\nu}}(q) a_{\mu}(q), \quad
\Bb_{\mu\nu}(q) = \adag_{\mu}(q) a_{\nu}(q). 
\end{align}
Explicit expressions for the matrix elements, 
$N^{(\tau)}_A, N^{(\tau)}_B, F_{A,s}^{(\pm)}$ and $F_{B,s}^{(\pm)}$,  
are given in Appendix \ref{app:quasiparticle}.

We define the monopole--pairing gaps $\Delta_{0}^{(\tau)}(q)$, 
the quadrupole--pairing gaps $\Delta_{2,K=0,2}^{(\tau)}(q)$,
and the quadrupole deformations $D^{(+)}_{2,K=0,2}(q)$ by
\begin{align}
 \Delta_{0}^{(\tau)}(q) =&
 G_0^{(\tau)}\bra{\phi(q)}\Ahat^{(\tau)(+)}\ket{\phi(q)} \\
 \Delta_{2,K=0,2}^{(\tau)}(q) =&
 G_{2,K=0,2}^{(\tau)}\bra{\phi(q)}\Bhat^{(\tau)(+)}_{2,K=0,2(+)}\ket{\phi(q)},
 \\
 D^{(+)}_{2,K=0,2}(q) =& \bra{\phi(q)}\Dhat^{(+)}_{2,K=0,2}\ket{\phi(q)}.
\end{align}

\subsection{The moving-frame QRPA equations}
\label{sec:mfQRPA}

The infinitesimal generators, $\Qhat(q)$ and $\Phat(q)$, are represented
in the quasiparticle representation as
\begin{align}
 \Qhat(q) =&  Q^A(q) + Q^B(q) \nonumber \\
=& 
{\sum_{\mu\bar{\nu}}}' Q^A_{\mu\bar{\nu}}(q) 
( \Abdag_{\mu\bar{\nu}}(q) + \Ab_{\mu\bar{\nu}}(q))
+ {\sum_{\mu\nu}}' Q^B_{\mu\nu}(q) 
(\Bb_{\mu\nu}(q) + \Bb_{\bar{\mu}\bar{\nu}}(q)), \\
 \Phat(q) =& i{\sum_{\mu\bar{\nu}}}' P_{\mu\bar{\nu}}(q)
( \Abdag_{\mu\bar{\nu}}(q) - \Ab_{\mu\bar{\nu}}(q)).
\end{align}
In the following, we discuss how to get the $n$-th solution of the 
moving-frame QRPA equations in {\it Step 3} 
assuming that the $(n-1)$-th solution $\Qhat^{(n-1)}(q)$ is already known.
For later convenience, we introduce the following one-body operator
\begin{align}
 \Rhat^{(\pm)}_s =& [\Fhatpm_s,\frac{\del V}{\del q} \Qhat^{(n-1)}(q)]
 \nonumber \\
=&
 \bra{\phi(q)}\Rhat^{(\pm)}_s\ket{\phi(q)} +
 {\sum_{\mu\bar{\nu}}}' R^{(\pm)}_{A,s}(\mu\bar{\nu})(\Abdag_{\mu\bar{\nu}}
 \mp \Ab_{\mu\bar{\nu}})
 + {\sum_{\mu\nu}}' R^{(\pm)}_{B,s}(\mu\nu)(\Bb_{\mu\nu} +
 \Bb_{\bar{\mu}\bar{\nu}}),
\end{align}
with
\begin{align}
 R^{(\pm)}_{A,s}(\mu\bar{\nu}) = \frac{\del V}{\del q}
{\sum_\rho}'  & \left( 
 F_{B,s}^{(\pm)}(\mu\rho)(Q^{A}_{\rho\bar{\nu}})^{(n-1)} \pm
 (Q^{A}_{\mu\bar{\rho}})^{(n-1)} F_{B,s}^{(\pm)}(\bar{\rho}\bar{\nu}) \right.
 \nonumber \\
& \left. - (Q^{B}_{\mu\rho})^{(n-1)}   F_{A,s}^{(\pm)}(\rho\bar{\nu}) 
- F_{A,s}^{(\pm)}(\mu\bar{\rho}) (Q^{B}_{\bar{\rho}\bar{\nu}})^{(n-1)}  \right).
\end{align}
We can express the matrix elements, 
$Q^{A}_{\mu\bar{\nu}}(q)$ and $P_{\mu\bar{\nu}}(q)$, 
using Eqs. (\ref{eq:ascc2_sep}) and (\ref{eq:ascc3_sep}) as
\begin{align}
 (Q^{A}_{\mu\bar{\nu}}(q))^{(n)} =& 
 {\sum_{\mu'\bar{\nu'}}}' g_2(\mu\bar{\nu},\mu'\bar{\nu'}) \left\{ 
 \sum_s \left( \Fp_{A,s}(\mu'\bar{\nu'}) \fp_{PR,s}(q) - 
 \Rm_{A,s}(\mu'\bar{\nu'})\fm_{Q,s}(q) \right) \right. \nonumber \\
 &\left. + \sum_{\tau} N^{(\tau)}(\mu'\bar{\nu'}) f_{N}^{(\tau)}(q) \right\}
 + g_1(\mu\bar{\nu},\mu'\bar{\nu'}) \sum_s \Fm_{A,s}(\mu'\bar{\nu'}) 
 \fm_{Q,s}(q), \label{eq:Qeq-pqq}\\
 P_{\mu\bar{\nu}}(q) =&
 {\sum_{\mu'\bar{\nu'}}}' g_3(\mu\bar{\nu},\mu'\bar{\nu'}) \left\{
 \sum_s \left( \Fp_{A,s}(\mu'\bar{\nu'}) \fp_{PR,s}(q) -
 \Rm_{A,s}({\mu'\bar{\nu'}})\fm_{Q,s}(q) \right) \right. \nonumber \\
 &\left. + \sum_{\tau} N^{(\tau)}(\mu'\bar{\nu'}) f_{N}^{(\tau)}(q) \right\}
 + g_4(\mu\bar{\nu},\mu'\bar{\nu'}) \sum_s \Fm_{A,s}(\mu'\bar{\nu'})
 \fm_{Q,s}(q), \label{eq:Peq-pqq}
\end{align}
where $\fp_{PR,s}(q)\equiv \fp_{P,s}(q) + \fp_{R,s}(q)$. 
The metrics $g_i, (i=1\sim 4)$ are defined by
\begin{align}
 g_1(\mu\bar{\nu},\mu'\bar{\nu'}) \equiv &
 (\Mc^{-1} \Ec )_{\mu\bar{\nu},\mu'\bar{\nu'}}, &\quad
 g_2(\mu\bar{\nu},\mu'\bar{\nu'}) \equiv & 
 (\Mc^{-1})_{\mu\bar{\nu}, \mu'\bar{\nu'}}, \\
 g_3(\mu\bar{\nu},\mu'\bar{\nu'}) \equiv & 
 (\Ec \Mc^{-1})_{\mu\bar{\nu}, \mu'\bar{\nu'}}, & \quad
 g_4(\mu\bar{\nu},\mu'\bar{\nu'}) \equiv &  (\Ec\Mc^{-1}\Ec)_{\mu\bar{\nu},\mu'\bar{\nu'}} - \delta_{\mu\mu'}\delta_{\bar{\nu}\bar{\nu'}},
\end{align}
where $\Mc$ and $\Ec$ are given by
\begin{align}
 \Mc_{\mu\bar{\nu},\mu'\bar{\nu'}}(\omega^2(q)) =& \{(E_{\mu}+E_{\bar{\nu}})^2 -
 \omega^2(q)\}\delta_{\mu\mu'}\delta_{\bar{\nu}\bar{\nu'}}
 \nonumber \\
 &+ \delta_{\mu\mu'} \left(\frac{1}{2}E_{\mu'} + E_{\nu'} -
 \frac{1}{2}E_{\bar{\nu}}\right) (Q^{B}_{\bar{\nu'}\bar{\nu}})^{(n-1)} \frac{\del
 V}{\del q}(q),
\nonumber \\
 & + (Q^{B}_{\mu\mu'})^{(n-1)}\left(
 E_{\mu'} - \frac{1}{2}E_{\mu} + \frac{1}{2}E_{\bar{\nu'}}
\right) \delta_{\bar{\nu}\bar{\nu'}} \frac{\del V}{\del q}(q),
\\
 \Ec_{\mu\bar{\nu},\mu'\bar{\nu'}} = & (E_{\mu} + E_{\bar{\nu}})
 \delta_{\mu\mu'} \delta_{\bar{\nu}\bar{\nu'}}.
\end{align}
The quantities (\ref{eq:f-mfQRPA}) and 
the canonical variable condition (\ref{eq:can-np}) are 
expressed in terms of $(\Qhat^{A}(q))^{(n)}$ and $\Phat(q)$ as
\begin{align}
 \fm_{Q,s}(q) =& -\kappa_s
 \bra{\phi(q)}[\Fhatm_s,\Qhat^{(n)}(q)]\ket{\phi(q)}
 = -2\kappa_s(\Fm_{A,s}, (Q^A(q))^{(n)}), \label{eq:f_Q}\\
 \fp_{PR,s}(q)=& \kappa_s
 \bra{\phi(q)}[\Fhatp_s,\frac{1}{i}B(q)\Phat(q)]\ket{\phi(q)} \nonumber
 \\
 & - \frac{1}{2}
 \kappa_s \bra{\phi(q)}[[\Fhatp_s, \frac{\del V}{\del
 q}\Qhat^{(n-1)}(q)],
 \Qhat^{(n)}(q)]\ket{\phi(q)} \nonumber  \\
 =& 2\kappa_s (\Fp_{A,s}, P(q)) - \kappa_s (\Rp_{A,s}, (Q^A(q))^{(n)}),
 \label{eq:f_PR}
\end{align}
\begin{align}
 \bra{\phi(q)}[\Nhat^{(\tau)},\Phat(q)]\ket{\phi(q)} = -2i(N^{(\tau)}_A,P(q)) = 0,
 \label{eq:f_N}
\end{align}
where 
\begin{align}
 (X,Y) \equiv {\sum_{\mu\bar{\nu}}}' X(\mu\bar{\nu}) Y(\mu\bar{\nu}).
\end{align}
Substituting Eqs.~(\ref{eq:f_Q}), (\ref{eq:f_PR}), (\ref{eq:f_N}) into
Eqs.~(\ref{eq:Qeq-pqq}) and (\ref{eq:Peq-pqq}), we derive the 
dispersion equation
\begin{align}
\bm{S}\cdot\bm{f} =
\sum_{s'\tau'}
\begin{pmatrix}
 S^{Q,Q}_{ss'} && S^{Q,PR}_{ss'} && S^{Q,N}_{s\tau'} \\ \\
 S^{PR,Q}_{ss'}&& S^{PR,PR}_{ss'}&& S^{PR,N}_{s\tau'} \\ \\
 S^{N,Q}_{\tau s'} && S^{N,PR}_{\tau s'} && S^{N,N}_{\tau\tau'}
\end{pmatrix}
\begin{pmatrix}
 \fm_{Q,s'}(q) \\ \\ \fp_{PR,s'}(q) \\ \\ f_{N}^{(\tau')}(q)
\end{pmatrix}
= 0, \label{eq:disp}
\end{align}
where the matrix elements of $\bm{S}$ are given by
\begin{subequations}
\begin{align}
 S^{Q,Q}_{ss'} =&
 2 (\Fm_{A,s}, \Fm_{A,s'})_{g_1}
-2 (\Fm_{A,s}, \Rm_{A,s'})_{g_2} -\frac{1}{\kappa_s}\delta_{ss'},\\
 S^{Q,PR}_{ss'}=&
 2 (\Fm_{A,s},\Fp_{A,s'})_{g_2}, \\
 S^{Q,N}_{s\tau'} =&
 2 (\Fm_{A,s}, N_A^{(\tau')})_{g_2}, \\
 S^{PR,Q}_{ss'} =&
 2(\Fp_{A,s},\Fm_{A,s'})_{g_4}
-2(\Fp_{A,s},\Rm_{A,s'})_{g_3} \nonumber \\
&+(\Rp_{A,s},\Fm_{A,s'})_{g_1}
 -(\Rp_{A,s},\Rm_{A,s'})_{g_2}, \\
 S^{PR,PR}_{ss'} =&
 2 (\Fp_{A,s},\Fp_{A,s'})_{g_3}
 + (\Rp_{A,s},\Fp_{A,s'})_{g_2}
 -\frac{1}{\kappa_s} \delta_{ss'},  \\
 S^{PR,N}_{s\tau'} =&
 2 (\Fp_{A,s}, N_A^{(\tau')})_{g_3} +
 (\Rp_{A,s},N_A^{(\tau')})_{g_2}, \\
 S^{N,Q}_{\tau s'} =&
 (N_A^{(\tau)},\Fm_{A,s'})_{g_4} -(N_A^{(\tau)},\Rm_{A,s'})_{g_3}, \\
 S^{N,PR}_{\tau s'} =& (N_A^{(\tau)}, \Fp_{A,s'})_{g_3}, \\
 S^{N,N}_{\tau\tau'} =& (N_A^{(\tau)}, N_A^{(\tau')})_{g_3}.
\end{align}
\end{subequations}
The parentheses in the above matrix elements are defined by
\begin{align}
 (X,Y)_{g_i} = {\sum_{\mu\bar{\nu}\mu'\bar{\nu'}}}'
X({\mu\bar{\nu}}) g_i({\mu\bar{\nu}\mu'\bar{\nu'}}) Y({\mu'\bar{\nu'}}),
\ \ \ (i = 1\sim 4).
\end{align}
As we mentioned before, the ASCC equations are invariant 
under the gauge transformation associate with number fluctuations. 
The quantities,  
\begin{align}
\fm_{Q,s=1,2}(q) = -2G_0^{(\tau=n,p)} \bra{\phi(q)}
[\Ahat^{(\tau=n,p)(-)},\Qhat(q)]\ket{\phi(q)},
\end{align}
and $f_{N}^{(\tau)}(q)$, are transformed by (\ref{eq:transform}) as
\begin{align}
 \fm_{Q,s=1,2}(q) \rightarrow & \fm_{Q,s=1,2}(q) - 4\alpha^{(\tau=n,p)} \Delta_{0}^{(\tau=n,p)}(q),
 \\
 f_N^{(\tau=n,p)}(q) \rightarrow & f_N^{(\tau=n,p)}(q) - \alpha^{(\tau=n,p)} \omega^2(q).
\end{align}
Thus, we have to fix the gauge in solving the dispersion equation 
(\ref{eq:disp}). 
For both neutrons and protons, 
we choose the ``ETOP'' gauge\cite{hin07}
\begin{align}
 \fm_{Q,s=1}(q) = 0, \quad \fm_{Q,s=2}(q) = 0.
\end{align}
This gauge fixing condition reduces 
the dimension of the dispersion equations. 
We can then use the submatrix $\bm{S'}$ of $\bm{S}$, 
where terms related to the anti-Hermite part of the monopole pairing 
operators, ($\Fhatm_{s=1,2}$), are dropped.
From Eq.~(\ref{eq:disp}), the moving-frame QRPA frequency squared $\omega^2(q)$ is determined
by the condition
\begin{align}
 \det \bm{S'}(\omega^2(q)) = 0.
\end{align}
The lowest $\omega^2(q)$ solution (including negative values)
are considered as the most collective mode at $q$.
Note that we consider imaginary $\omega(q)$ solutions as well as 
real ones. Once $\omega^2(q)$ is determined, $\bm{f}(q)$,
$(Q^{A}_{\mu\bar{\nu}}(q))^{(n)}$ and $P_{\mu\bar{\nu}}(q)$ are
obtained by use of the normalization condition 
\begin{align}
 \bra{\phi(q)}[(\Qhat^{A}(q))^{(n)}, \Phat(q)]\ket{\phi(q)} =
2i ((Q^{A}(q))^{(n)}, P(q)) = i.
\end{align}

\section{Requantization of the collective Hamiltonian}
\label{sec:quantum}
\subsection{Requantization and construction of wave functions 
in the laboratory frame}
\label{subsec:requantization}

Solving the basic equations of the ASCC method and Thouless--Valatin equations,
we obtain the collective Hamiltonian (\ref{eq:Hcoll});
we can put the collective mass $B(q)^{-1}$ to unity without losing generality 
because it just defines the scale of measuring the length of the collective
path. We also put the number fluctuation $\bm{n}$ to zero. 
Requantization is done simply by replacing the classical variables  
with the quantum operators:
\begin{align}
 p \rightarrow \frac{\hbar}{i}\frac{\del}{\del q}, \quad
I_i \rightarrow \Ihat_i.
\end{align}
The Schr\"odinger equation for the requantized collective
Hamiltonian is 
\begin{align}
 \left( -\frac{1}{2}\frac{\del^2}{\del q^2} + \sum_{i=1}^3 \frac
 {\hat{I}_i^2} {2\Jc_i(q)}+ V(q) \right) \Psi_{IMk}(q,\Omega)
= E_{I,k} \Psi_{IMk}(q,\Omega).
\label{eq:schrodinger}
\end{align}
The collective wave function in the laboratory frame, 
$\Psi_{IM,k}(q, \Omega)$, is a function of 
the collective coordinate $q$ and three Euler angles $\Omega$,  
and specified by the total angular momentum $I$, 
its projection $M$ on the laboratory $z$-axis, 
and the index $k$ distinguishing different quantum states 
having the same $I$ and $M$. Note that the three components $\hat{I}_i$ 
of the angular momentum oprerator are defined with respect to 
the principal axes $(1,2,3)\equiv(x',y',z')$ associate with 
the moving-frame HB state $\ket{\phi(q)}$. 

Using the rotational wave functions ${\Dc}^I_{MK}(\Omega)$, 
we can write the collective wave functions in the laboratory frame as 
\begin{align}
 \Psi_{IMk}(q,\Omega) =&
\sum_{K=-I}^I \Phi'_{IKk}(q) \sqrt{\frac{2I+1}{8\pi^2}}
 {\Dc}^{I}_{MK}(\Omega) \label{eq:wavefunc1} \\
 =& \sum_{K=0}^I \Phi_{IKk}(q) \product{\Omega}{IMK}. \label{eq:wavefunc2} 
\end{align}
Here $\Phi'_{IKk}$ are intrinsic wave functions that represent 
large-amplitude collective vibrations responsible for 
the oblate--prolate shape mixing. They are specified, instead of $M$, by 
the projection $K$ of the angular momentum on the intrinsic $z'$-axis. 
We assume that the intrinsic states have the positive signature. 
Then, their $K$ and $-K$ components are connected by
\begin{align}
 \Phi_{IKk}'(q) = (-)^I \Phi_{I-Kk}'(q).
\end{align}
Accordingly, it is convenient to use new rotational wave functions  
defined by 
\begin{align}
 \product{\Omega}{IMK} = \frac{1}{\sqrt{2(1+\delta_{K0})}}
 \sqrt{\frac{2I+1}{8\pi^2}}
 \left(
 {\Dc}^{I}_{MK}(\Omega) + (-)^{I}{\Dc}^{I}_{M-K}(\Omega)
\right),
\end{align}
and new vibrational wave functions 
\begin{align}
 \Phi_{IKk}(q) = \sqrt{\frac{2}{1+\delta_{K0}}}\Phi'_{IKk}(q) = (-)^I
 \sqrt{\frac{2}{1+\delta_{K0}}}\Phi'_{I-Kk}(q),
\end{align}
in place of $\Phi'_{IK,k}$.
As the $\Dc$ functions are normalized as
\begin{align}
 \int d\Omega {\Dc}^{I \ast}_{MK}(\Omega) {\Dc}^{I'}_{M'K'}(\Omega) =
 \frac{8\pi^2}{2I+1} \delta_{II'} \delta_{MM'} \delta_{KK'},
\end{align}
the normalization of the vibrational wave functions is given by  
\begin{align}
 \int dq \sum_{K=0}^I \Phi^{\ast}_{IKk}(q) \Phi_{IKk'}(q) = \delta_{kk'}.
 \label{eq:normalization} 
\end{align}

\subsection{Boundary Conditions}

Multiplying a rotational wave function $\product{\Omega}{IMK}$ to 
the Schr\"odinger equation (\ref{eq:schrodinger}) from the left 
and integrating out the Euler angles $\Omega$, we obtain 
the collective Schr\"odinger equation for large-amplitude vibration:
\begin{align}
 \left(- \frac{1}{2}\frac{\del^2}{\del q^2} + V(q)
 \right) \Phi_{IKk}(q)
 + \sum_{K'=0}^I
 \bra{IMK}\hat{T}_{\rm rot}\ket{IMK'} \Phi_{IK'k}(q)  = E_{I,k}
 \Phi_{IKk}(q),
 \label{eq:shrodinger2}
\end{align}
where $\hat{T}_{\rm rot}=\sum_i {\hat{I}_i^2}/(2\Jc_i(q))$. 

The boundary conditions can be specified by projecting the collective path 
to the $(\beta, \gamma)$ plane and by using the well-known symmetry properties 
of the Bohr-Mottelson's collective Hamiltonian.~\cite{BMtextII,kum67}. 
The deformation parameters $\beta$ and $\gamma$ are defined by
\begin{align}
 \beta(q)\cos\gamma(q) =& \chi'
 \bra{\phi(q)}\Dhat^{(+)}_{20}\ket{\phi(q)}
/ (\hbar\omega_0b^2), \label{eq:def-bg1} \\
 \beta(q)\sin\gamma(q) =& \sqrt{2}\chi'
\bra{\phi(q)}\Dhat^{(+)}_{22}\ket{\phi(q)}
/ (\hbar\omega_0 b^2). \label{eq:def-bg2}
\end{align}
and measure the magnitude and triaxiality of quadrupole deformation 
of the HB mean-field in the moving-frame as functions of the collective coordinate $q$. 
Here, $\hbar\omega_0$ denotes the frequency of the harmonic--oscillator 
potential, $\chi'\equiv\chi b^4$, and the harmonic--oscillator length parameter 
$b$ is related to the radius parameter $r_0$ by
\begin{align}
 b^2 = \frac{4}{5} \left(\frac{2}{3}\right)^{\frac{1}{3}}
r_0^2 A^{\frac{1}{3}}.
\end{align}

The boundary conditions for the vibrational collective
wave functions depend on the character of the collective path in the
$(\beta, \gamma)$ plane. 
As we will discuss later in \S\ref{sec:68Se-path},
the collective path for $^{68}$Se passes through the $\gamma$-direction 
(see Fig.\ref{fig:68Se-path}). In this case, the following boundary conditions 
are employed at the prolate and oblate limits.
At the prolate limit, $\gamma(q_{\rm pro})=0^{\circ}$, 
the vibrational wave functions are required to satisfy
\begin{align}
\Phi_{IKk}(q_{\rm pro} - q) = (-)^{\frac{K}{2}} \Phi_{IKk}(q_{\rm pro}
+ q), \label{eq:periodic-boundary-pro}
\end{align}
which is equivalent to
\begin{align}
 \Phi_{IKk}(q_{\rm pro}) = 0  & \quad (K = 2,6,\cdots), \\
 \frac{d\Phi_{IKk}}{dq}\Big\arrowvert_{q=q_{\rm pro}} = 0 & \quad
 (K=0,4,\cdots).
\end{align}
At the oblate limit, $\gamma(q_{\rm ob})=60^{\circ}$, 
the HB mean-field is symmetric about the intrinsic $y'$-axis, and then 
the boundary conditions are given by~\cite{kum67}
\begin{align}
 \Phi_{IKk}(q_{\rm ob} - q) =  (-)^{\frac{K}{2}}\sum_{K'} 
\frac{2}{\sqrt{(1+\delta_{K0}) (1 + \delta_{K'0}) }}
\Dc^{I}_{KK'}(\frac{\pi}{2},\frac{\pi}{2},\pi)
\Phi_{IK'k}(q_{\rm ob} + q). \label{eq:periodic-boundary-ob}
\end{align}

%\begin{align}
% \Phi_{IKk}(q_{\rm ob} - q) =  \sum_{K'} M^I_{KK'}
% \Phi_{IK'k}(q_{\rm ob} + q), \label{eq:periodic-boundary-ob}
%\end{align}
%where
%\begin{align}
% M^I_{KK'} =& (-)^{\frac{K}{2}} \frac{2}{
%\left[(1+\delta_{K0}) (1 + \delta_{K'0}) \right]^{\frac{1}{2}}}
%\Dc^{I}_{KK'}(\frac{\pi}{2},\frac{\pi}{2},\pi) \nonumber \\
% =& \frac{2^{-I+1}}{\left[
%                     (1+\delta_{K0})(1+\delta_{K'0})
%                    \right] ^{\frac{1}{2}}}
%\sum_S \frac{(-)^{I-S}\left[
%                        (I+K)!(I-K)!(I+K')!(I-K')!
%                       \right]^{\frac{1}{2}}}
%{S!(K+K'+S)!(I-K-S)!(I-K'-S)!}.
%\end{align}
In the case of $^{72}$Kr, the collective path connecting the oblate and
prolate shapes are not periodic with respect to the $\gamma$-direction
(see Fig.~\ref{fig:72Kr-path}). Accordingly, we set the box boundary conditions 
at the edge of the path:  
\begin{align}
 \Phi_{IKk}(q_{\rm min}) = \Phi_{IKk}(q_{\rm max}) = 0. 
 \label{eq:box-boundary}
\end{align}

The matrix elements $\bra{IMK}\hat{T}_{\rm rot}\ket{IMK'}$
of the rotational kinetic energy operator in Eq. (\ref{eq:shrodinger2}) 
can be easily calculated:
\begin{align}
 \bra{IMK}\hat{T}_{\rm rot}\ket{IMK} =& a(q)I(I+1) + b(q)K^2, \\
 \bra{IMK}\hat{T}_{\rm rot}\ket{IM,K+2} =& \bra{IM,K+2}\hat{T}_{\rm
 rot}\ket{IMK} \nonumber \\
 =& c(q) \{(I+K+2)(I+K+1)(I-K)(I-K-1)\}^{-\frac{1}{2}},
\label{eq:rotation0}
\end{align}
where 
\begin{align}
 a(q) =& \frac{1}{4}\left(\frac{1}{\Jc_1(q)} +
 \frac{1}{\Jc_2(q)}\right), 
\label{eq:rotation1} \\
 b(q) =& \frac{1}{4}\left(\frac{2}{\Jc_3(q)} 
        - \frac{1}{\Jc_1(q)} - \frac{1}{\Jc_2(q)}\right), 
\label{eq:rotation2} \\
 c(q) =& \frac{1}{8}\left(
 \frac{1}{\Jc_1(q)} - \frac{1}{\Jc_2(q)}
\right).
\label{eq:rotation3}
\end{align}
The other matrix elements are zero.

\subsection{Electric quadrupole moments and transitions}

To evaluate electric quadrupole (E2) moments and transition probabilities, 
we need to derive expressions of the E2 operator in the collective subspace. 
This can be easily achieved by the same procedure as we have derived the 
quantum collective Hamiltonian.  As described below, we first take expectation 
values of the E2 operators with respect to the moving-frame HB state $\ket{\phi(q,p)}$
and then apply the canonical quantization procedure.  

In accordance with the quadrupole operators (\ref{eq:quadrupole}), 
we define the E2 operators in the model space under consideration as
\begin{align}
 \Dhat^{\prime({\rm E2})}_\mu =& \sum_{\tau} e^{(\tau)}_{\rm eff}
 \sum_{kl\in\tau}
 D^{(\tau)}_{2\mu}(kl) \cdag_k c_l, \label{eq:defE2}\\
 {\Dhat}^{\prime({\rm E2})}_{\mu+} =& 
 \frac{1}{2}(\Dhat^{\prime({\rm E2})}_\mu +\Dhat^{\prime({\rm E2})}_{-\mu}),
\end{align}
where $e^{(\tau)}_{\rm eff}$ are effective charges. 
Their expectation values in the collective subspace are expanded 
up to second order in the collective momentum $p$ as 
\begin{align}
 D^{\prime({\rm E2})}_{\mu+}(q,p) 
= \bra{\phi(q,p)} {\Dhat}^{\prime({\rm E2})}_{\mu+}\ket{\phi(q,p)} \\
= D^{\prime({\rm E2})}_{\mu+}(q) + 
\frac{1}{2}D^{\prime\prime({\rm E2})}_{\mu+}(q) p^2,
\label{eq:E2coll}
\end{align}
where 
\begin{align}
D^{\prime({\rm E2})}_{\mu+}(q) =& 
\bra{\phi(q)} {\Dhat}^{\prime({\rm E2})}_{\mu+} \ket{\phi(q)}, \\
D^{\prime\prime({\rm E2})}_{\mu+}(q) =& -\bra{\phi(q)}
[[{\Dhat}^{\prime({\rm E2})}_{\mu+}, \Qhat(q)], \Qhat(q)] \ket{\phi(q)}.
\end{align}
The quantities $D^{\prime({\rm E2})}_{\mu+}(q,p)$ are called 
collective representations of the E2 operators.
Note that these are defined in the intrinsic frame associated with the 
moving-frame HB mean-field. 
We now apply the canonical quantization to them. 
Then, the collective coordinate $q$ and the collective momentum $p$ 
become quantum operators acting on vibrational wave functions 
$\Phi_{IKk}(q)$. We call the requantized E2 operators 
``collective E2 operators'' and denote them 
$\Dhat^{\prime({\rm E2})}_{\mu+}$. 
Thus, the E2 matrix elements  between two collective vibrational states 
are evaluated as
\begin{align}
\bra{\Phi_{IKk}} \Dhat^{\prime({\rm E2})}_{\mu+} \ket{\Phi_{IK'k'}}
=
\int dq \, \Phi_{IKk}^{\ast}(q) \left(
 D^{\prime({\rm E2})}_{\mu+}(q) - \frac{1}{2} \frac{d}{dq}
 D^{\prime\prime({\rm E2})}_{\mu+}(q) \frac{d}{dq}\right) \Phi_{IK'k'}(q).
\end{align}
We need to calculate these integrals only for vibrational states
which satisfies the selection rules of the E2 operators. 

The collective E2 operators $\Dhat^{\prime({\rm E2})}_{\mu}$ are defined 
in the intrinsic frame, and those in the laboratory frame 
$\Dhat^{({\rm E2})}_{\mu}$ are obtained by 
\begin{align}
 \Dhat^{({\rm E2})}_{\mu} = \sum_{\mu} \Dc^2_{\mu\mu'}(\Omega) 
 \Dhat^{\prime({\rm E2})}_{\mu\prime}.
\end{align}
As is well known, $B$(E2) values and spectroscopic quadrupole moments $Q(Ik)$
are given in terms of reduced matrix elements 
$\brared{Ik}\Dhat^{({\rm E2})}_{+}\ketred{Ik}$ as
\begin{align}
 B({\rm E2}; Ik \rightarrow I'k') = (2I+1)^{-1}\left| \brared{Ik}
 \Dhat^{({\rm E2})} \ketred{I'k'} \right|^2, \label{eq:BE2def}
\end{align}
\begin{align}
Q(Ik) &= \sqrt{\frac{16\pi}{5}} \bra{I,M=I,k}
\Dhat^{({\rm E2})} \ket{I,M=I,k}
 \nonumber \\
&= \sqrt{ \frac{16\pi}{5}}
\begin{pmatrix}
 I & 2 & I \\ -I & 0 & I
\end{pmatrix}
\brared{Ik}\Dhat^{({\rm E2})}_{\mu}\ketred{Ik}.
\end{align}
These reduced matrix elements can be evaluated by using 
the Wigner-Eckart theorem 
\begin{align}
 \bra{I,M=I,k} \Dhat^{({\rm E2})}_{0} \ket{I',M=I,k'} =
\begin{pmatrix}
 I & 2 & I' \\ -I & 0 & I
\end{pmatrix}
 \brared{I, k} \Dhat^{({\rm E2})} \ketred{I',{k'}}, 
 \label{eq:E2-wig}
\end{align}
and calculating the left-hand side as~\cite{kum67}
\begin{align}
\bra{I,M=I,k} & \Dhat^{({\rm E2})}_{0} \ket{I',M=I,k'}
\nonumber \\
=& \frac{\sqrt{(2I+1)(2I'+1)}}{8\pi^2}
\sum_{KK'\mu}
\bra{\Phi'_{IKk}}\Dhat^{\prime({\rm E2})}_{\mu}\ket{\Phi'_{I'K'k'}}
\bra{\Dc^I_{IK }}\Dc^2_{0\mu}\ket{\Dc^{I'}_{IK'}} 
\nonumber \\
=& \sqrt{(2I+1)(2I'+1)}
\sum_{KK'\mu} 
\bra{\Phi'_{IKk}}\Dhat^{\prime({\rm E2})}_{\mu}\ket{\Phi'_{I'K'k'}} 
\nonumber \\
 & (-)^{I-K} 
\begin{pmatrix}
I & 2 & I' \\ -I & 0 & I
\end{pmatrix}
\begin{pmatrix}
 I & 2 & I' \\ -K & \mu & K'
\end{pmatrix}.
\end{align}
In the intrinsic frame, the $\mu=\pm 1$ components of the collective 
E2 operator vanish and those for the $\mu=\pm 2$ components are equal. 
Thus we obtain 
\begin{align}
 \brared{Ik} & \Dhat^{({\rm E2})} 
  \ketred{I'k'} = \nonumber \\
 =& \sqrt{(2I+1)(2I'+1)} (-)^I
 \sum_{K\ge 0}
\left[
\begin{pmatrix}
I & 2 & I' \\ -K & 0 & K
\end{pmatrix}
\bra{\Phi_{IKk}}\Dhat^{\prime({\rm E2})}_{0+}\ket{\Phi_{I'K'k'}}
\right. \nonumber \\
&+ \sqrt{1+\delta_{K0}}
\left\{
\begin{pmatrix}
I & 2 & I' \\ -K-2 & 2 & K
\end{pmatrix}
\bra{\Phi_{I,K+2,k}}\Dhat^{\prime({\rm E2})}_{2+}\ket{\Phi_{I'Kk'}}
\right. \nonumber \\
& \left. \left. +
\begin{pmatrix}
I & 2 & I' \\ K & 2 & -K-2
\end{pmatrix}
(-)^{I+I'}
\bra{\Phi_{IKk}}\Dhat^{\prime({\rm E2})}_{2+} \ket{\Phi_{I',K+2,k'}}
\right\} \right]. \label{eq:BE2-final}
\end{align}

\section{Results of numerical calculation and discussions} 
\label{sec:result}

\subsection{Details of numerical calculation}

In numerical calculations,
we consider two major shells ($N_{\rm sh} = 3, 4$) for protons and neutrons 
and use the same values for the single-particle energies, the monopole pairing 
strength $G_0^{(\tau)}$ and the quadrupole particle-hole interaction strength 
$\chi$ as in Ref.~\citen{kob04}. 
The single-particle energies are listed in Table \ref{table:sp-energy}.
The interaction strengths were adjusted to approximately 
reproduce the pairing gaps and the quadrupole deformations obtained 
by the Skyrme-HFB calculation carried out by Yamagami et al.\cite{yam01}
These values are $G_0^{(n)} = G_0^{(p)}=0.320$ and 
$\chi'\equiv\chi b^4=0.248$ MeV for $^{68}$Se; 
$G_0^{(n)}=0.299, G_0^{(p)}=0.309$ and $\chi'=0.255$ MeV for $^{72}$Kr. 
The oscillator frequency and the radius parameters are set 
to $\hbar\omega_0=41.2A^{1/3}$ MeV and $r_0=1.2$ fm.
For the quadrupole pairing strength, we use the 
self-consistent value derived by Sakamoto and Kishimoto,~\cite{sak90}
\begin{align}
 G_{2K}^{(\tau){\rm self}} =  \left[
\sum_{\alpha\beta\in\tau} \frac{1}{4}
\left(
\frac{1}{E_{\alpha}} + \frac{1}{E_{\beta}}
\right)
 |D^{(\tau)}_{2K}(\alpha\beta)|^2
%|\bra{\alpha}\Qhat_{2K}\ket{\beta}|^2
\right]^{-1},
\end{align}
where $E_{\alpha}$ is the quasiparticle energy evaluated by 
the BCS approximation at the spherical shape. Accordingly,  
$G_{20}^{\rm self}=G_{21}^{\rm self}=G_{22}^{\rm self}$.

The effective charges $e^{(\tau)}_{\rm eff}$ are written as 
$e^{(n)}_{\rm eff} = \delta e_{\rm pol}$  for neutrons and 
$e^{(p)}_{\rm eff}= e + \delta e_{\rm pol}$ for protons. 
For simplicity, we use the same polarization charge 
$\delta e_{\rm pol}=0.905e$ 
for protons and neutrons, which is chosen to reproduce the experimental 
$B$(E2;$2^+_1\rightarrow 0^+_1$) value\cite{gad05} in $^{72}$Kr. 
Only this data is available for E2 transitions among low-lying states 
in $^{68}$Se and $^{72}$Kr. This value of $\delta e_{\rm pol}$ seems 
a little too large and needs further investigation.   
We take into account the momentum dependent term 
in the collective representation of the E2 operators, Eq.~(\ref{eq:E2coll}), 
although numerical calculations indicate that it gives only 
a few percent correction, at most, to the main term. 

In the present calculation, we ignore the curvature terms
(the fourth, fifth and sixth term in Eq.~(\ref{eq:ascc3_sep})),
to reduce the computational costs. We have checked that their contributions 
are negligible. 

In numerical calculation, careful treatment is necessary for the prolate 
limit, as the moment of inertia about the symmetry axis, $\Jc_3(q)$,
vanishes there. Actually, this does not cause a problem, because the 
$K \ne 0$ components of the vibrational wave function also vanish there.
To avoid numerical instability, however, we put  
$\Jc_3(q)=10^{-13} \hbar^2$ (MeV)$^{-1}$, for the prolate limit, 
and confirmed that this recipe works well without losing numerical accuracy.   
We applied this recipe also for the oblate limit where $\Jc_2(q)$  vanishes. 
Actually, the $y'$-axis component of the vibrational wave function 
also vanishes there, although it is not directly seen 
in Eq.~(\ref{eq:shrodinger2}) in which the wave functions are decomposed 
according to the $K$ quantum numbers choosing the $z'$-axis 
as the quantization axis.

\begin{table}[htbp]
\begin{center}
\caption{Energies in units of MeV of the spherical single-particle levels 
used in the calculation. These values are taken from Ref.~\citen{kob04}.}
\label{table:sp-energy}
\begin{tabular}{cccccccccc}\hline\hline
      orbits & $1f_{7/2}$ & $2p_{3/2}$ & $1f_{5/2}$ & $2p_{1/2}$ 
& $1g_{9/2}$ & $2d_{5/2}$ & $1g_{7/2}$ & $3s_{1/2}$ & $2d_{3/2}$ \\ \hline
neutrons & -9.02 & -4.93 & -2.66 & -2.21 & 0.00 & 5.27 & 6.36 & 8.34 &
 8.80 \\
protons & -8.77 & -4.23 & -2.41 & -1.50 & 0.00 & 6.55 & 5.90 & 10.10 &
 9.83 \\ \hline
\end{tabular}
\end{center}
\end{table}

\subsection{Properties of local minima in $^{68}$Se and $^{72}$Kr}
We summarize in Table \ref{table:HB} the results of calculation 
for the properties of the HB equilibrium states (local minima 
in the potential energy surface).
For both $^{68}$Se and $^{72}$Kr, the lowest HB minimum possesses an oblate 
shape, while the second minimum is prolate.
The energy differences between the oblate and prolate minima  
evaluated using the P+Q Hamiltonian with (without) the quadrupole
pairing interaction are 300 keV (196 keV) for $^{68}$Se
and 827 keV (626 keV) for $^{72}$Kr.
We find no qualitative change in the mean-field properties 
due to the inclusion of the quadrupole pairing interaction.

\begin{table}
\begin{center}
\caption{The quadrupole deformations and the pairing gaps 
$\Delta_0^{(\tau)}$ (in MeV) and $\Delta_{2K}^{(\tau)}$ (in MeV fm$^2$)
at the HB local minima in $^{68}$Se and $^{72}$Kr, calculated with the P+Q 
 Hamiltonian including the quadrupole pairing interaction.} \label{table:HB}
 \begin{tabular}{ccccccccc} \hline\hline
  ($G_2=G_2^{\rm self}$) & $\beta$ & $\gamma$ & $\Delta_{0}^{(n)}$ & $\Delta_0^{(p)}$ &
  $\Delta_{20}^{(n)}$ & $\Delta_{20}^{(p)}$ & $\Delta_{22}^{(n)}$ &
  $\Delta_{22}^{(p)}$ \\ \hline
$^{68}$Se (oblate) & 0.30 & 60$^{\circ}$ & 1.17 & 1.26 & 0.08 & 0.09 &
  0.10 & 0.11 \\ 
$^{68}$Se (prolate)& 0.26 & 0$^{\circ}$  & 1.34 & 1.40 & 0.14 & 0.15 & 0
  & 0 \\ 
$^{72}$Kr (oblate) & 0.35 & 60$^{\circ}$ & 0.92 & 1.06 & 0.05 & 0.06 &
  0.06 & 0.07 \\
$^{72}$Kr (prolate)& 0.38 & 0$^{\circ}$ & 1.14 & 1.27 & 0.19 & 0.19 & 0
  & 0  \\ \hline
\end{tabular}
\end{center}
\end{table}

The QRPA collective modes at the oblate and prolate minima can be classified 
in terms of the projections of angular momenta 
on the symmetry axis, $K_y$ and $K_z \equiv K$, respectively. 
Table \ref{table:QRPA} summarizes the properties the QRPA collective modes
at the oblate and prolate minima.
In $^{68}$Se, the lowest modes are $\gamma$-vibrational ($K_y$ or $K_z$=2) 
and the second lowest modes are $\beta$-vibrational ($K_y$ or $K_z$=0)
both at the oblate and the prolate minima.
It is seen that the quadrupole pairing interaction lowers their 
excitation energies without changing their ordering. 
In $^{72}$Kr, the lowest QRPA modes at the two minima are both 
$\beta$-vibrational, if the quadrupole pairing interaction is ignored.
Note, however, that the $K_z=0$ and 2 modes at the prolate second minimum
are close in energy and their ordering changes when the quadrupole pairing 
interaction is taken into account, whereas the lowest mode at the lowest 
oblate minimum is always $\beta$-vibrational.

\begin{table}[htbp]
\begin{center}
\caption{The excitation energies $\omega$ (in MeV) and the $K$ quantum numbers 
of the lowest two QRPA modes at the oblate and prolate minima 
in $^{68}$Se and $^{72}$Kr. The results of the calculation 
with ($G_2=G^{\rm self}_2$) and without ($G_2 = 0$)
including the quadrupole pairing interaction are compared.
The $K$ quantum numbers here represent $K_y$ or $K_z$ 
according to the shape (oblate or prolate).   
}
\label{table:QRPA}
\begin{tabular}{ccccccccc} \hline\hline
 & $G_2 = 0$ &  &          &   & $G_2=G^{\rm self}_2$ & & &  \\
 & $\omega_1$ & $K_1$ & $\omega_2$ & $K_2$ & $\omega_1$ & $K_1$ & $\omega_2$ & $K_2$ \\ \hline 
 $^{68}$Se (oblate) & 1.555 & 2 & 2.342 & 0 & 1.373 & 2 & 2.131 & 0 \\
 $^{68}$Se (prolate)& 1.015 & 2 & 1.915 & 0 & 0.898 & 2 & 1.369 & 0 \\
 $^{72}$Kr (oblate) & 1.150 & 0 & 1.909 & 0 & 1.239 & 0 & 2.010 & 2 \\
 $^{72}$Kr (prolate)& 1.606 & 0 & 1.674 & 2 & 1.644 & 2 & 1.714 & 0 \\ \hline
\end{tabular}
\end{center}
\end{table} 

\begin{figure}[tbhp]
\begin{center}
\begin{tabular}{c}
\includegraphics[width=100mm]{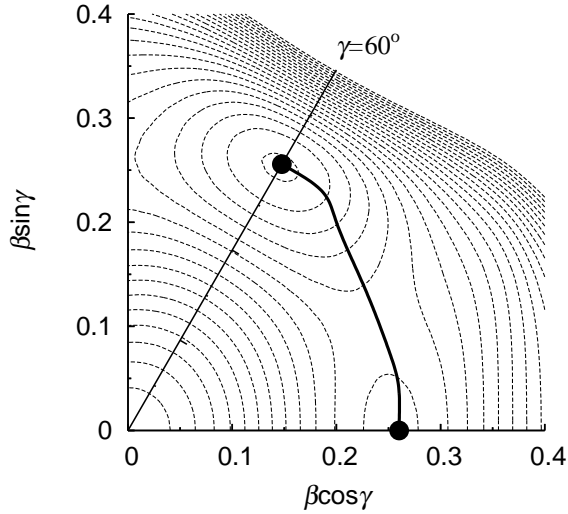}
\end{tabular}
\end{center}
\caption{The collective path for $^{68}$Se calculated with the P+Q
Hamiltonian including the quadrupole pairing interaction.
The path is projected onto the $(\beta, \gamma)$ potential energy
 surface.
The dots in the figure indicates the HB local minima.
The equipotential lines are drawn every 100 keV.}
\label{fig:68Se-path}
\end{figure}
\begin{figure}[htbp]
\begin{center}
\begin{tabular}{c}
\includegraphics[width=135mm]{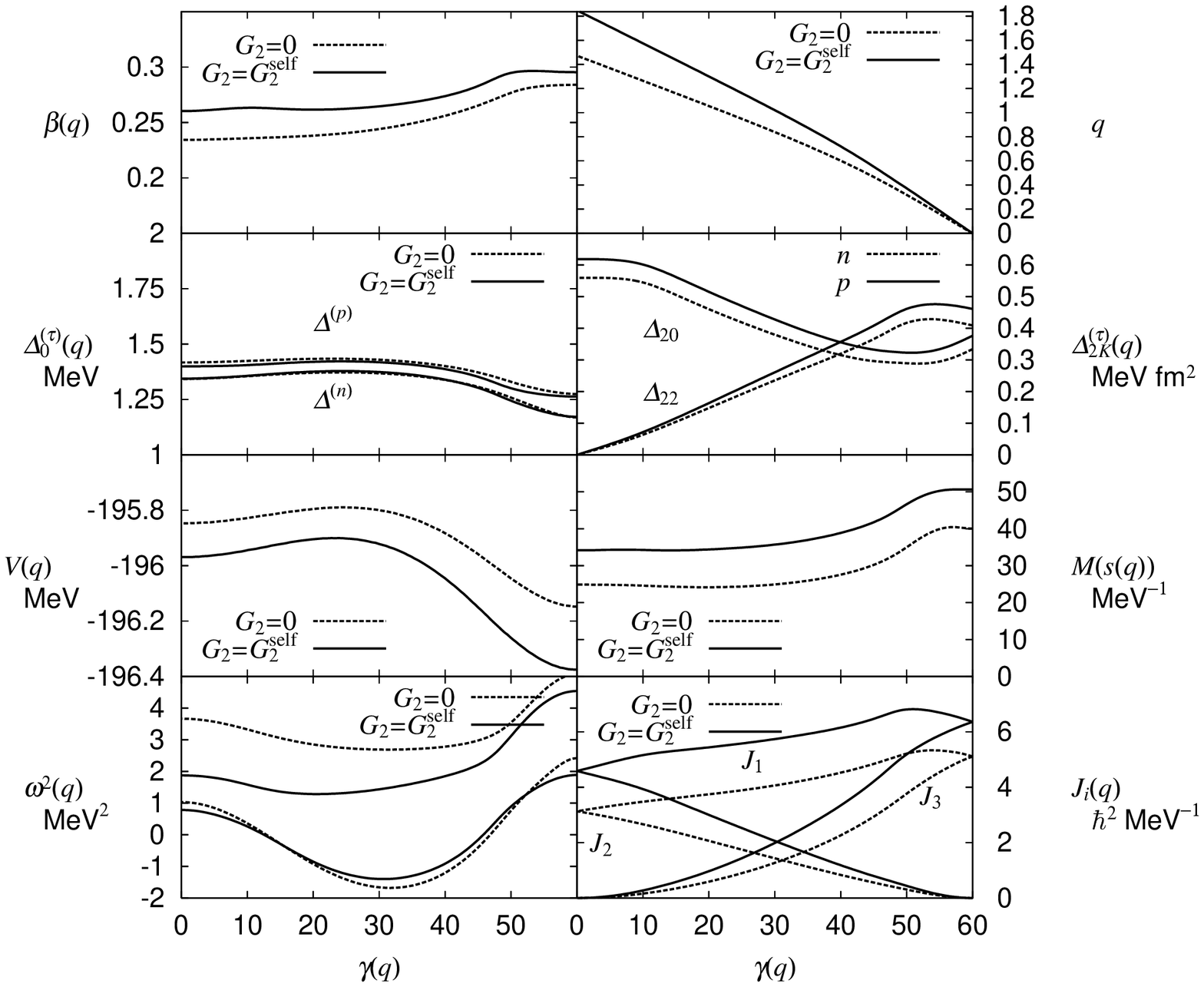}
\end{tabular}
\end{center}
\caption{Results of the calculation for $^{68}$Se. The monopole pairing
 gap
$\Delta_{0}^{(\tau)}(q)$, the quadrupole pairing gaps
$\Delta_{20}^{(\tau)}(q)$ and $\Delta_{22}^{(\tau)}(q)$,
the collective potential $V(q)$, the collective mass $M(s(q))$,
the rotational moments of inertia $\Jc_i(q)$,
the lowest two moving-frame QRPA frequencies squared $\omega^2(q)$,
the axial quadrupole deformation $\beta(q)$ and the canonical
collective coordinate $q$ are plotted as functions of $\gamma(q)$.
Results of the two calculations using the P+Q Hamiltonian with
($G_2=G_2^{\rm self}$) and without ($G_2=0$) the quadrupole pairing
interaction are compared.}
\label{fig:68Se}
\end{figure}

\subsection{Collective path connecting the oblate and prolate minima 
in $^{68}$Se} \label{sec:68Se-path}

We start by solving the basic equations of the ASCC method from the oblate
minimum ($q=0$) and progressively determine the collective path 
following the algorithm outlined in \S\ref{subsec:algorithm}.
Figure~\ref{fig:68Se-path} illustrates the collective path thus obtained 
by projecting it onto the $(\beta, \gamma)$ potential energy surface.
The path connects the two local minima passing through a potential valley
lying in the triaxial deformed region. 
The collective path in $^{68}$Se obtained with the P+Q Hamiltonian 
including the quadrupole pairing interaction is very similar to that 
obtained in Ref.~\citen{kob04} in which its effect was ignored.
As solutions of the ASCC equations we obtain various quantities; 
the canonical collective coordinate $q$, 
the quadrupole deformations, $\beta(q)$ and $\gamma(q)$,
the monopole and quadruple pairing gaps $\Delta_0^{(\tau)}(q)$ and 
$\Delta_{2K}^{(\tau)}(q)$, 
the collective potential $V(q)$, the collective mass $M(s(q))$,
the moving-frame QRPA frequency squared $\omega^2(q)$, and
the three rotational moments of inertia $\Jc_i(q)$.
These quantities are plotted in Fig.~\ref{fig:68Se} as functions of $\gamma(q)$.
The quadrupole deformation $\beta(q)$ is almost constant along
the collective path while the triaxial deformation $\gamma(q)$ 
varies and changes the oblate shape to the prolate shape.
It is seen that the quadrupole pairing interaction slightly increases 
the $\beta(q)$ values for all values of $\gamma(q)$.

The collective mass $M(s(q))$ plotted in Fig.~\ref{fig:68Se}
is defined as a function of the geometrical length, 
$ds=\sqrt{d\beta^2+\beta^2 d\gamma^2}$, in the $(\beta,\gamma)$ plane: 
\begin{align}
 M(s(q)) = M(q)/\{ (d\beta/dq)^2 +\beta^2(d\gamma/dq )^2 \}.
\end{align}
As explained in \S\ref{subsec:requantization}, 
we can put $M(q)=B(q)^{-1}=1 {\rm MeV}^{-1}$ here. 
We have found that the quadrupole pairing interaction increases 
the collective mass. This enhancement takes place almost independent of 
$\gamma(q)$, and is mainly due to the decrease of $d\gamma/dq$ along
the collective path.

Because the HB mean-field becomes symmetric about the $y'$- and $z'$-axes 
in the oblate and prolate limits, respectively, 
the rotational moment of inertia about the $y'$ ($z'$)-axis vanishes 
and the other two moments take the same values at the oblate (prolate) minimum. 
Their $\gamma$-dependence is similar to that of the irrotational moments of 
inertia. It is found the rotational moments of inertia increase about 
20$\sim$30$\%$ by the effect of the quadrupole pairing interaction.  
This enhancement as well as that of the inertial functions $M(s(q))$ is
due to the time-odd pair field generated by the quadrupole pairing interaction.

\subsection{Collective path connecting the oblate and prolate minima
in $^{72}Kr$} \label{sec:72Kr-path}

\begin{figure}[htbp]
\begin{center}
\begin{tabular}{c}
\includegraphics[width=100mm]{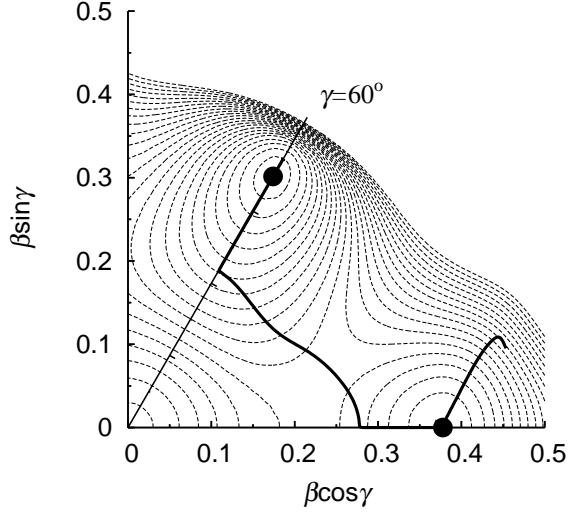}
\end{tabular}
\end{center}
\caption{The same as Fig.~\ref{fig:68Se-path} but for $^{72}$Kr.}
\label{fig:72Kr-path}
\end{figure}

\begin{figure}[htbp]
\begin{center}
\begin{tabular}{c}
\includegraphics[width=135mm]{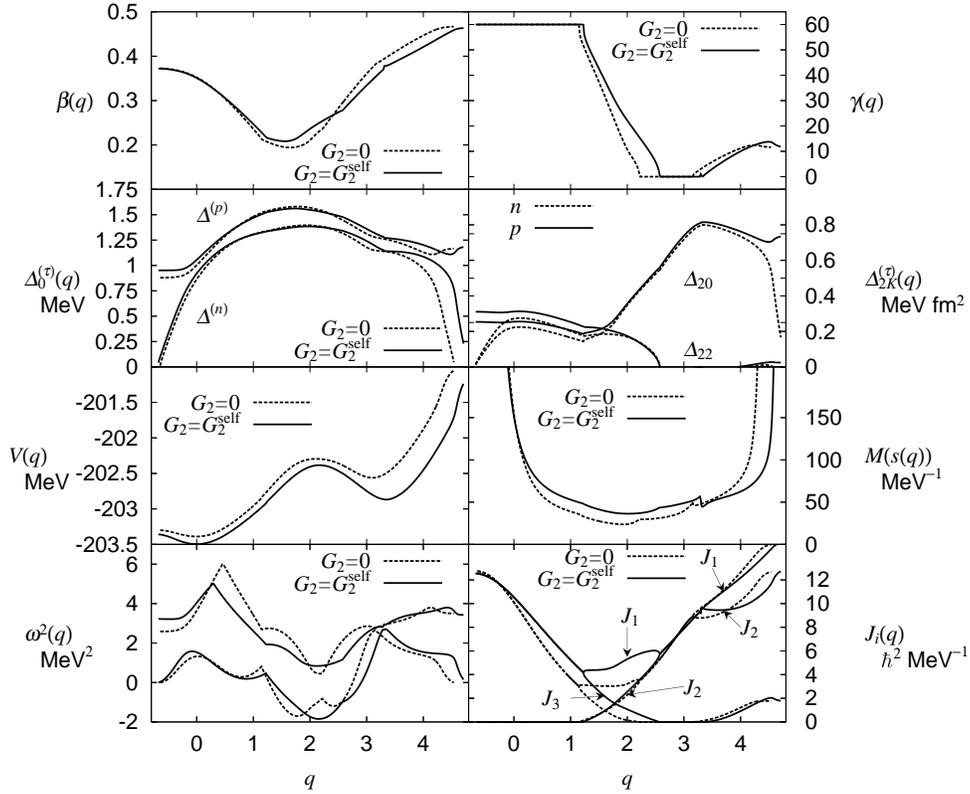}
\end{tabular}
\end{center}
\caption{
Results of the calculation for $^{72}$Kr.
Notations are the same as in Fig.~\ref{fig:68Se},
except that the quantities are plotted as functions of $q$
along the collective path.
The point $q=0$ corresponds to the oblate minimum,
while the prolate local minimum is located around $q=3.3$  ($q=3.1$)
for the calculation using the P+Q Hamiltonian with (without) the
quadrupole pairing interaction.}
\label{fig:72Kr}
\end{figure}

As for $^{68}$Se, we have determined the collective path for $^{72}$Kr 
starting from the oblate lowest minimum. 
The collective path projected onto the $(\beta, \gamma)$ plane is shown in 
Fig.~\ref{fig:72Kr-path}, and various quantities defined along the 
collective path are plotted in Fig.~\ref{fig:72Kr} as functions of $q$.
The collective paths calculated with and without including 
the quadrupole pairing interaction are similar.
Because the lowest mode of the moving-frame QRPA equations is 
$\beta$-vibrational around the oblate lowest minimum, 
the path at first goes along the axially symmetric line. 
Around $(\beta,\gamma)=(0.2,60^{\circ})$, the character of the lowest 
mode changes to $\gamma$-vibrational, and the path deviates from 
the axially symmetric line. 
When the collective path reaches the $\gamma=0^\circ$ line, 
the character of the lowest mode again changes to $\beta$-vibrational.
Approaching the prolate second minimum, 
the lowest mode changes its character once more 
to $\gamma$-vibrational and the collective path deviates from 
the $\gamma=0^\circ$ line. 

We note that the lowest two modes at the prolate second minimum are very 
close and their ordering in energy may be sensitive to the interactions used. 
We examined that, for example, the lowest mode at the prolate second minimum 
becomes $\beta$-vibrational, if the quadrupole pairing interaction is switched 
off, and in this case the axial symmetry breaking takes place at 
a larger $\beta$ value beyond the prolate second minimum.
In such a situation, two collective coordinates may be needed 
to describe the collective dynamics in a better way. 
This serves as an interesting subject for future investigation.
It should be emphasized that such a problem arises only locally 
in a small region on the $(\beta, \gamma)$ plane and 
the collective path is well defined globally. 

At large $\beta$ region beyond the oblate minima ($q<0$) 
along the $\gamma=60^\circ$ line, 
the lowest $K=0$ mode exhibits a strong mixture of $\beta$-vibration 
(fluctuation of axially symetric shape) and neutron pairing vibration
(fluctuation of pairing gaps), and the calculation to find the collective path 
eventually stops when the neutron monopole pairing collapses.

As for $^{68}$Se, we have found that the collective mass and 
the rotational moments of inertia increase also for $^{72}$Kr
due to the time-odd pair field generated by the quadrupole pairing interaction.
We note that the collective mass $M(s(q))$ diverges at large deformation. 
This behavior, seen also in the previous work\cite{kob03,kob04,hin06},
is associated with the disappearance of the pairing gaps. 

\subsection{Excitation spectra and quadrupole transitions 
in $^{68}$Se} \label{sec:68Se-energy}

\begin{figure}[htbp]
\begin{center}
\includegraphics[width=140mm]{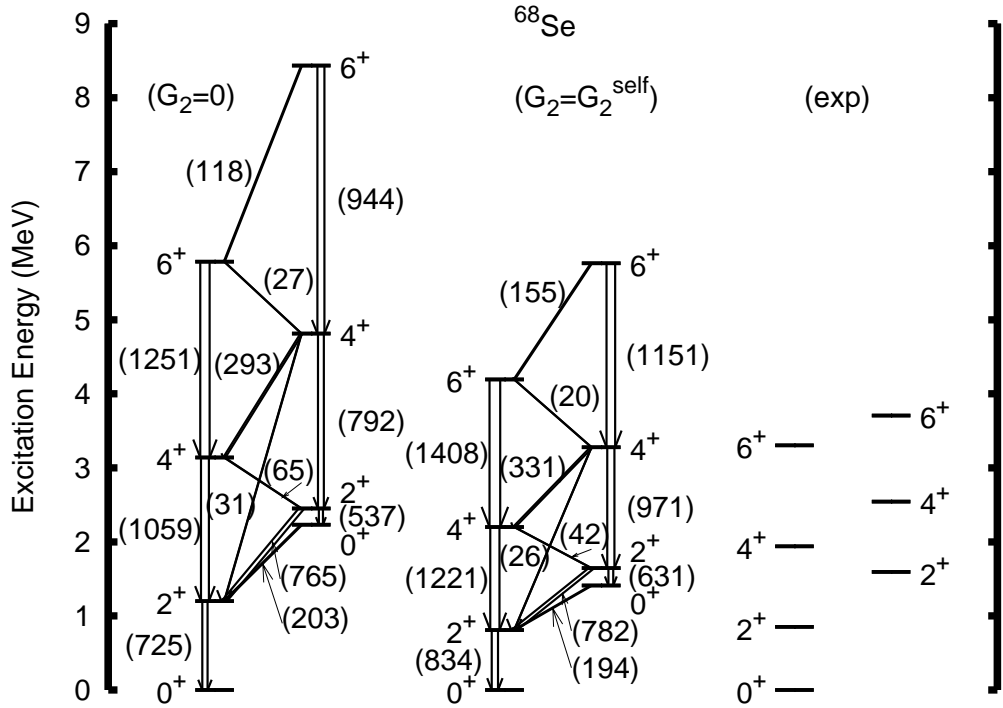}
\end{center}
\caption{
Excitation spectra and $B$(E2) values of low-lying states in $^{68}$Se
calculated by means of the ASCC method.
In the left (middle) panel, the quadrupole pairing is ignored (included)
in the microscopic Hamiltonian.
Experimental data\cite{fis00} are displayed in the right panel.
The $B$(E2) values larger than 1 W.u.
are indicated in parentheses besides the arrows in units of $e^2$
 fm$^4$.}
\label{fig:68Se-energy}
\end{figure}

\begin{figure}[htbp]
\begin{center}
\includegraphics[width=100mm]{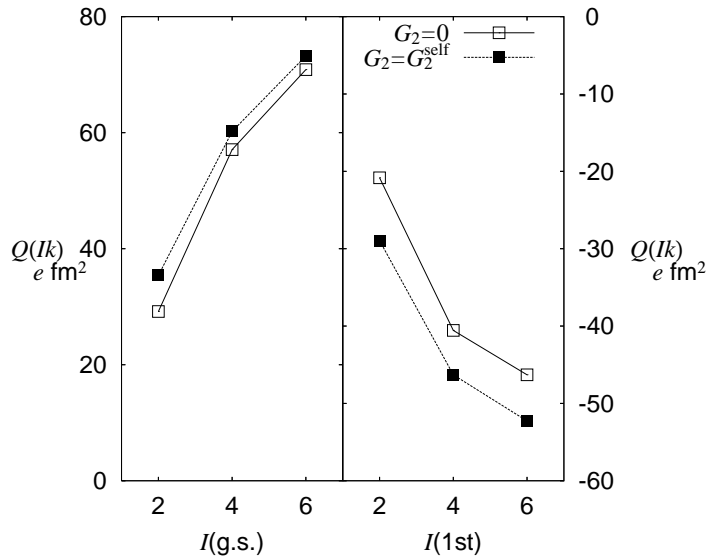}
\end{center}
\caption{
Spectroscopic quadrupole moments of low-lying states in $^{68}$Se.
The left and right panels show the spectroscopic
quadrupole moments of the yrast states and of the second lowest
states in each angular momentum, respectively.
The units for the right panels are indicated
besides the right vertical lines.
Results of calculation with (without) including the quadrupole pairing
interaction are indicated by filled (open) squares.}
\label{fig:68Se-Qsp}
\end{figure}

\begin{figure}[htbp]
\begin{center}
\includegraphics[width=120mm]{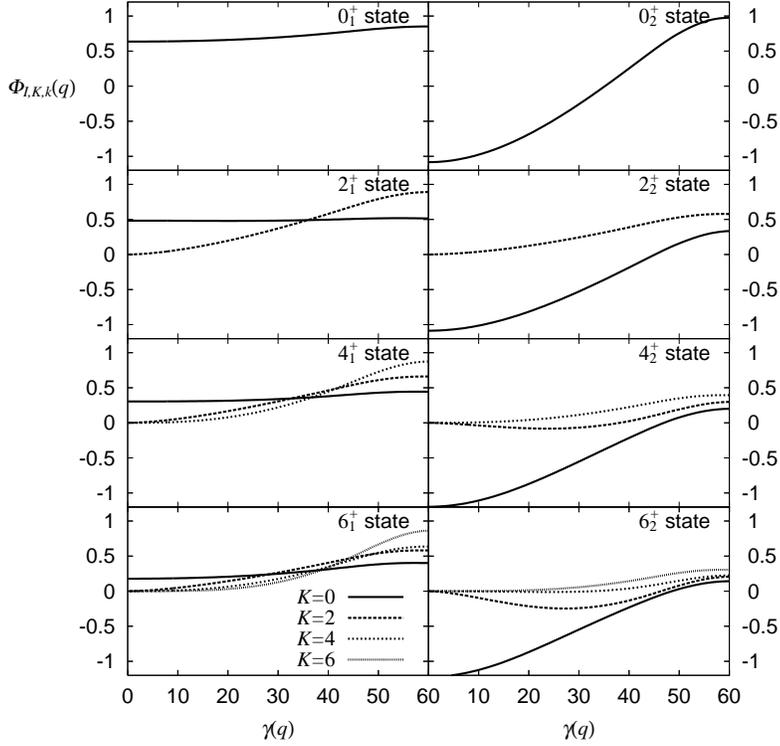}
\end{center}
\caption{Vibrational wave functions $\Phi_{IKk}(q)$ of
the yrast states ({\it left})  and
the second lowest states in each angular momentum ({\it right}):
in $^{68}$Se,
In each panel, different $K$ components of the
vibrational wave functions are plotted as functions of $\gamma(q)$.
The calculation is performed with the P+Q Hamiltonian
including the quadrupole pairing interaction.}
\label{fig:68Se-wave}
\end{figure}

\begin{figure}[htbp]
\begin{center}
\includegraphics[width=120mm]{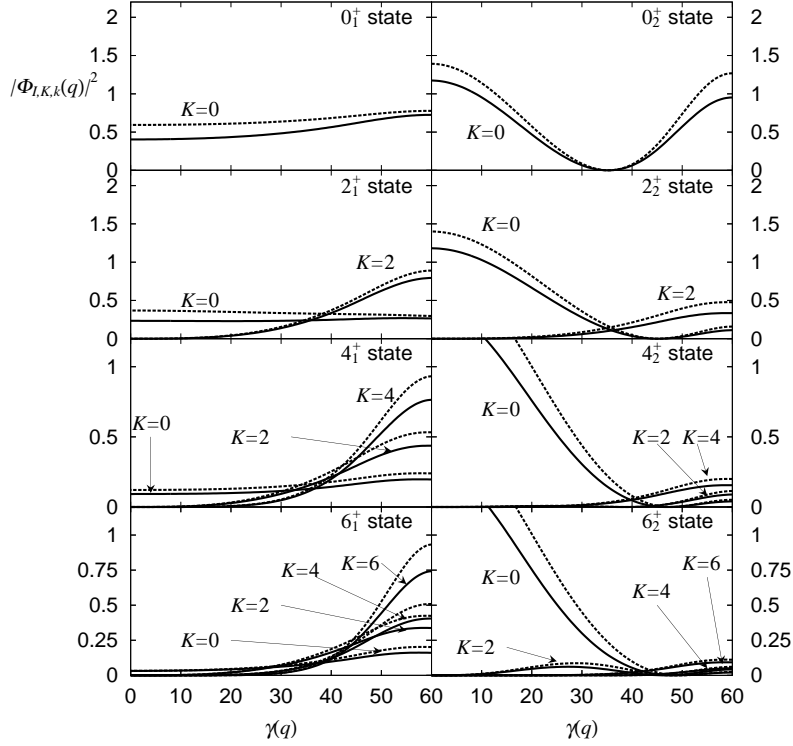}
\end{center}
\caption{
Vibrational wave functions squared $|\Phi_{IKk}(q)|^2$
the yrast states ({\it left})  and
the second lowest states in each angular momentum ({\it right}):
in $^{68}$Se,
In each panel, different $K$ components are plotted as functions of
 $\gamma(q)$.
The solid (dashed) line shows the result of calculation
using the P+Q Hamiltonian with (without) the quadrupole pairing
 interaction.
Note that the vibrational wave functions are normalized
as Eq.~(\ref{eq:normalization}) with respect to the collective
 coordinate $q$,
so that the factor $d\gamma/dq$ is multiplied to them when integrating
with respect the triaxial deformation parameter $\gamma$. The
 $d\gamma/dq$
values calculated including the quadrupole pairing interaction are
 larger
than those without including it for all range of $q$.
}
\label{fig:68Se-wavesquared}
\end{figure}

\begin{figure}[htbp]
\begin{center}
\begin{tabular}{c}
\includegraphics[width=100mm]{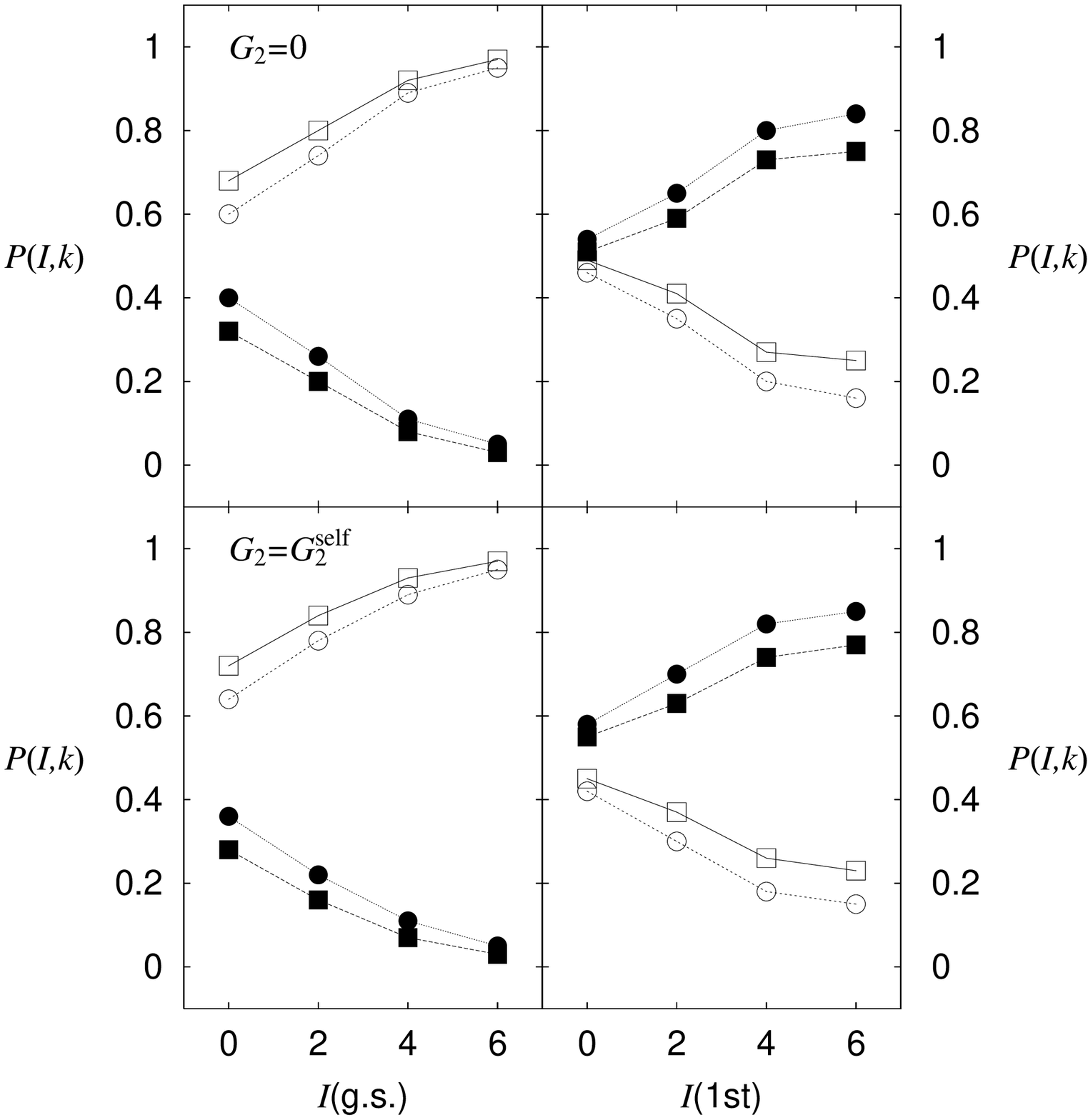}
\end{tabular}
\end{center}
\caption{
The oblate and prolate probabilities evaluated
for individual eigen-states in $^{68}$Se.
The upper (lower) panel shows the probabilities calculated using the P+Q
Hamiltonian without (with) the quadrupole pairing interaction.
The probabilities defined by setting the boundary at the barrier top
($\gamma=30^{\circ}$) are shown by squares (circles).}
\label{fig:68Se-mixing}
\end{figure}

The collective Schr\"odinger equation (\ref{eq:schrodinger}) is solved 
with the boundary conditions 
(\ref{eq:periodic-boundary-pro}) and (\ref{eq:periodic-boundary-ob})
for $^{68}$Se to get energy spectra, quadrupole moments and transition 
probabilities.
The result of calculation is displayed in Fig.~\ref{fig:68Se-energy}.
The calculation yields the excited prolate rotational bands as well as
the oblate ground state band.
It is seen that the inter-band E2 transitions are weaker than the
intra-band E2 transitions, indicating that the oblate--prolate 
shape coexistence picture holds.
The calculation suggests the existence of the excited $0^+$ state 
which has not yet been found in experiment.
The spectroscopic quadrupole moments presented in Fig.~\ref{fig:68Se-Qsp} are 
also consistent with the oblate--prolate shape coexistence picture:
The yrast states possess positive spectroscopic quadrupole moments 
indicating the oblate deformation,
while the second lowest states for each angular momentum have negative value
indicating the prolate deformation.
In Fig.~\ref{fig:68Se-energy}, the excitation spectra calculated with and 
without including the quadrupole pairing interaction are compared.
We see that the quadrupole pairing plays an important role in decreasing 
the excitation energies. 
This is because the time-odd pair field generated by the quadrupole pairing 
enhances the collective mass and the rotational moments of inertia.

In Fig.~\ref{fig:68Se-wave}, the vibrational wave functions are presented.
One may notice that the behaviors of the $0^+$ states are significantly 
different from the $I \ne 0$ states:   
The vibrational wave functions of the lowest and the second lowest $0^+$ 
states spread over the entire collective path indicating 
that the oblate and prolate shapes are strongly mixed
via the triaxial degree of freedom.
In contrast to the $0^+$ states, 
the $I \ne 0$ wave functions contain the $K\ne 0$ components 
which take the maximum values at the oblate limit. 
We can see this trend more clearly by plotting the collective wave functions 
squared. This is presented in Fig.~\ref{fig:68Se-wavesquared}. 
The vibrational wave function of the ground $0^+$ state spreads over 
the entire region of $\gamma$, while that of the excited  $0^+$ state 
exhibits prominent peaks both at the oblate and prolate limits. 
In contrast, the vibrational wave functions of the $I \ne 0$ yrast states 
are localized around the oblate shape, 
while those of the second lowest states (in each angular momentum)  
are localized around the prolate shape. 
The localization develops with increasing angular momentum. 
In the yrast states, all the $K \ne 0 $ components exhibit the maxima 
at the oblate shape, while the $K=0$ component dominates at the prolate shape 
in the second lowest states. 

In order to evaluate the oblate--prolate shape mixing in a more quantitative 
manner, we define the oblate and prolate probabilities as follows:
\begin{align}
 P_{\rm ob}(I,k) = \int_{q_{\rm min}}^{q_0} dq \sum_{K=0}^{I}
 |\Phi_{IKk}(q)|^2, \quad  
 P_{\rm pro}(I,k)= \int_{q_0}^{q_{\rm max}} dq \sum_{K=0}^{I} 
 |\Phi_{IKk}(q)|^2, 
\end{align}
where we assume that 
$q_{\rm min}\le q_{\rm ob} < q_0 < q_{\rm pro} \le q_{\rm max}$.
The ``boundary'' between the oblate and the prolate regions is set to 
the top of the potential barrier between the two minima, or at $\gamma=30^\circ$.
Figure~\ref{fig:68Se-mixing} shows these probabilities for $^{68}$Se.
The oblate and prolate states are strongly mixed in the $0^+$ states. 
It is clearly seen that the shape mixing rapidly decreases as the
angular momentum increases.

\subsection{Excitation spectra and quadrupole transitions in $^{72}$Kr}\label{sec:72Kr-energy}

\begin{figure}[htbp]
\begin{center}
\includegraphics[width=140mm]{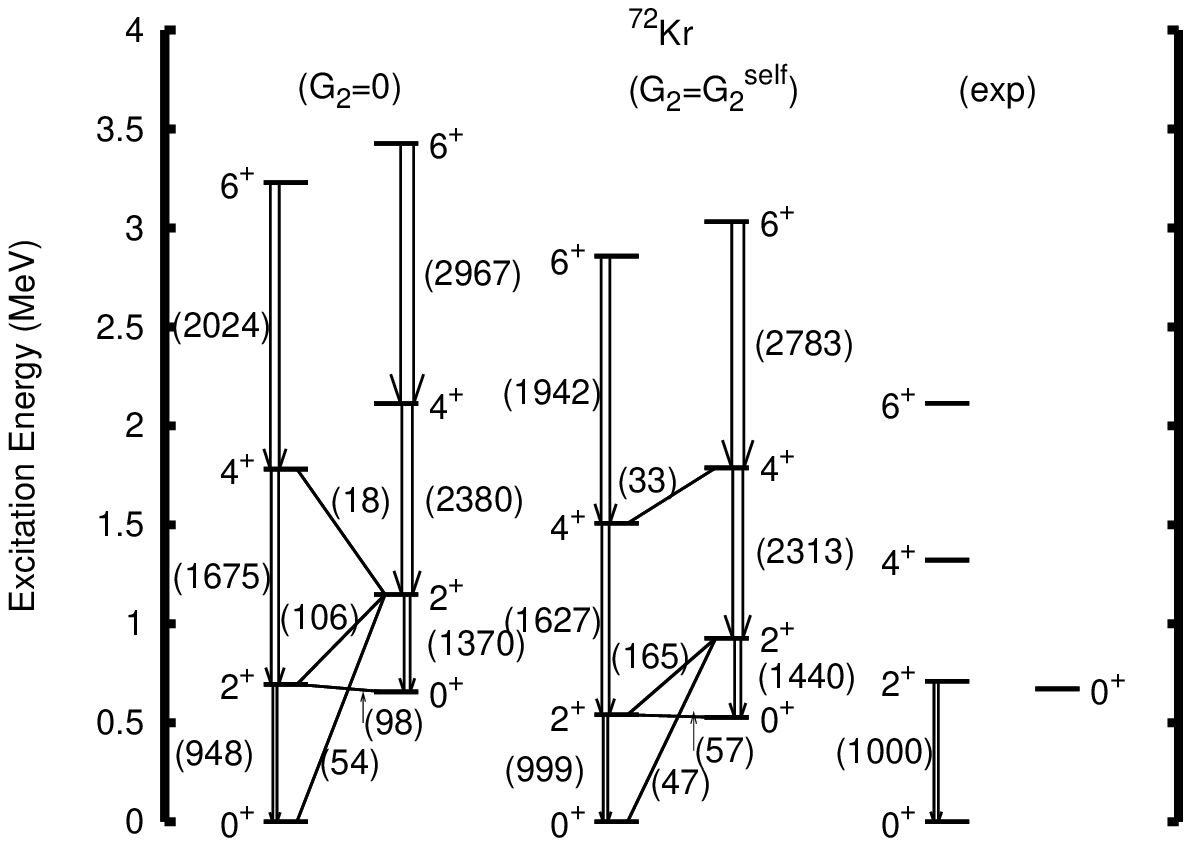}
\end{center}
\caption{
Same as Fig.~\ref{fig:68Se-energy} but for $^{72}$Kr.
Experimental data are taken from Refs.~\citen{fis03, bou03, gad05}.}
\label{fig:72Kr-energy}
\end{figure}

\begin{figure}[htbp]
\begin{center}
\includegraphics[width=100mm]{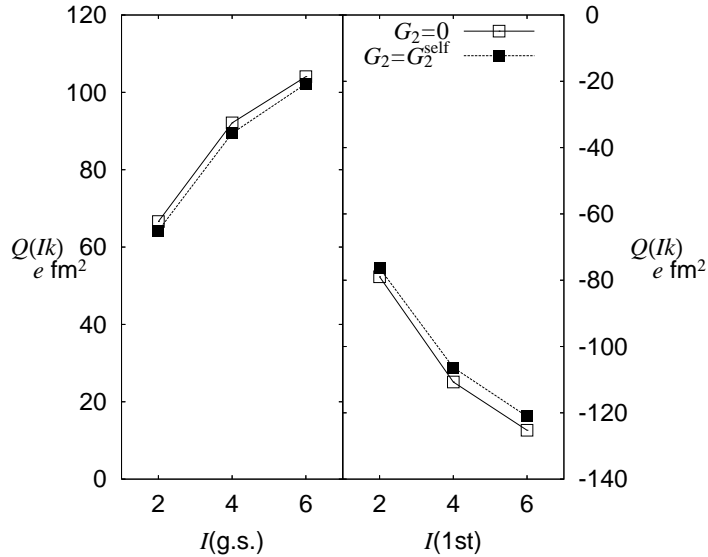}
\end{center}
\caption{
Same as Fig.~\ref{fig:68Se-Qsp} but for $^{72}$Kr.
See the caption of Fig.~\ref{fig:68Se-Qsp}}
\label{fig:72Kr-Qsp}
\end{figure}

\begin{figure}[htbp]
\begin{center}
\includegraphics[width=120mm]{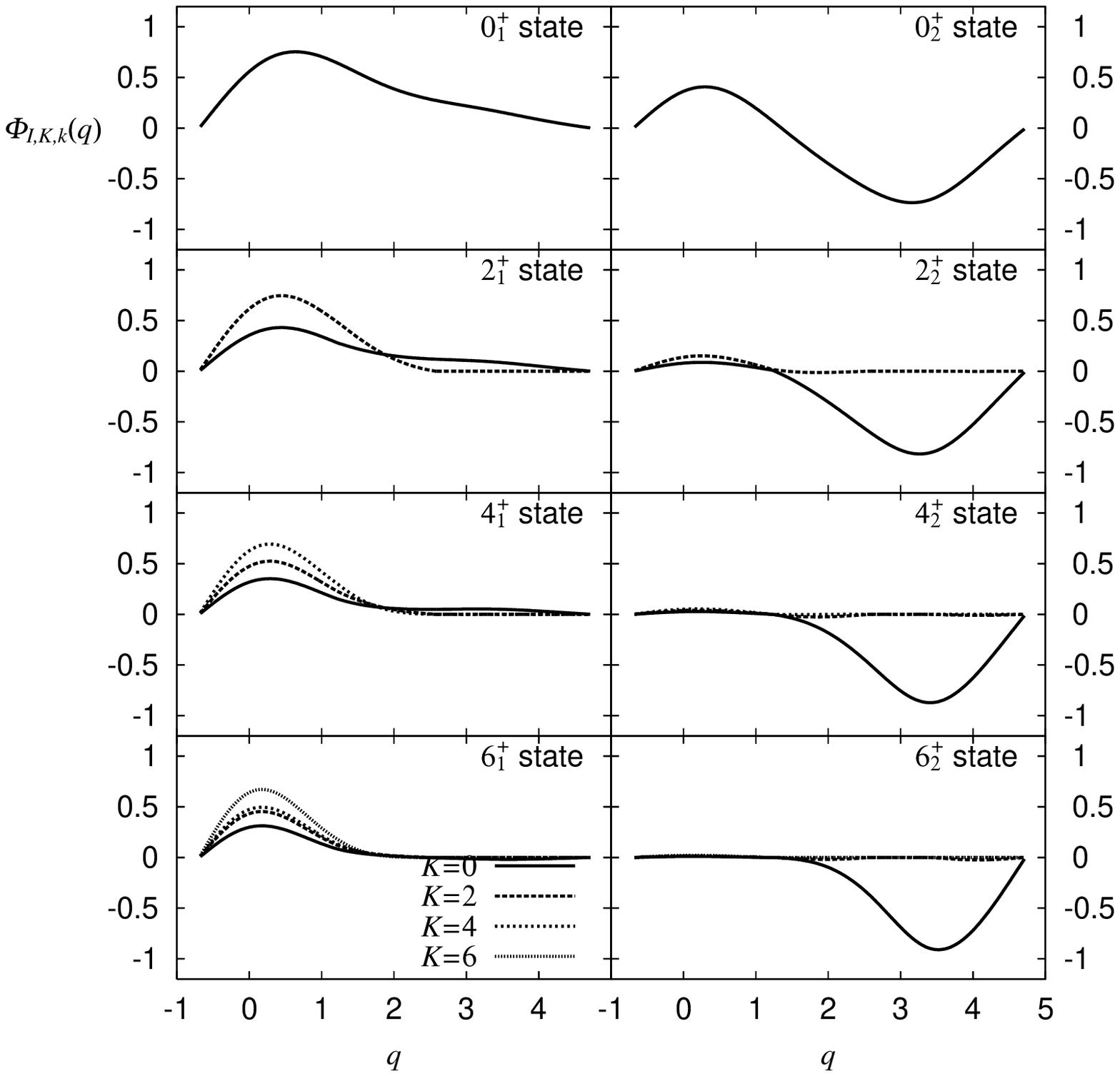}
\end{center}
\caption{Vibrational wave functions $\Phi_{IKk}(q)$ of low-lying states
 in
$^{72}$Kr plotted as functions of $q$.
See the caption to Fig.~\ref{fig:68Se-wave}.}
\label{fig:72Kr-wave}
\end{figure}

\begin{figure}[htbp]
\begin{center}
\includegraphics[width=120mm]{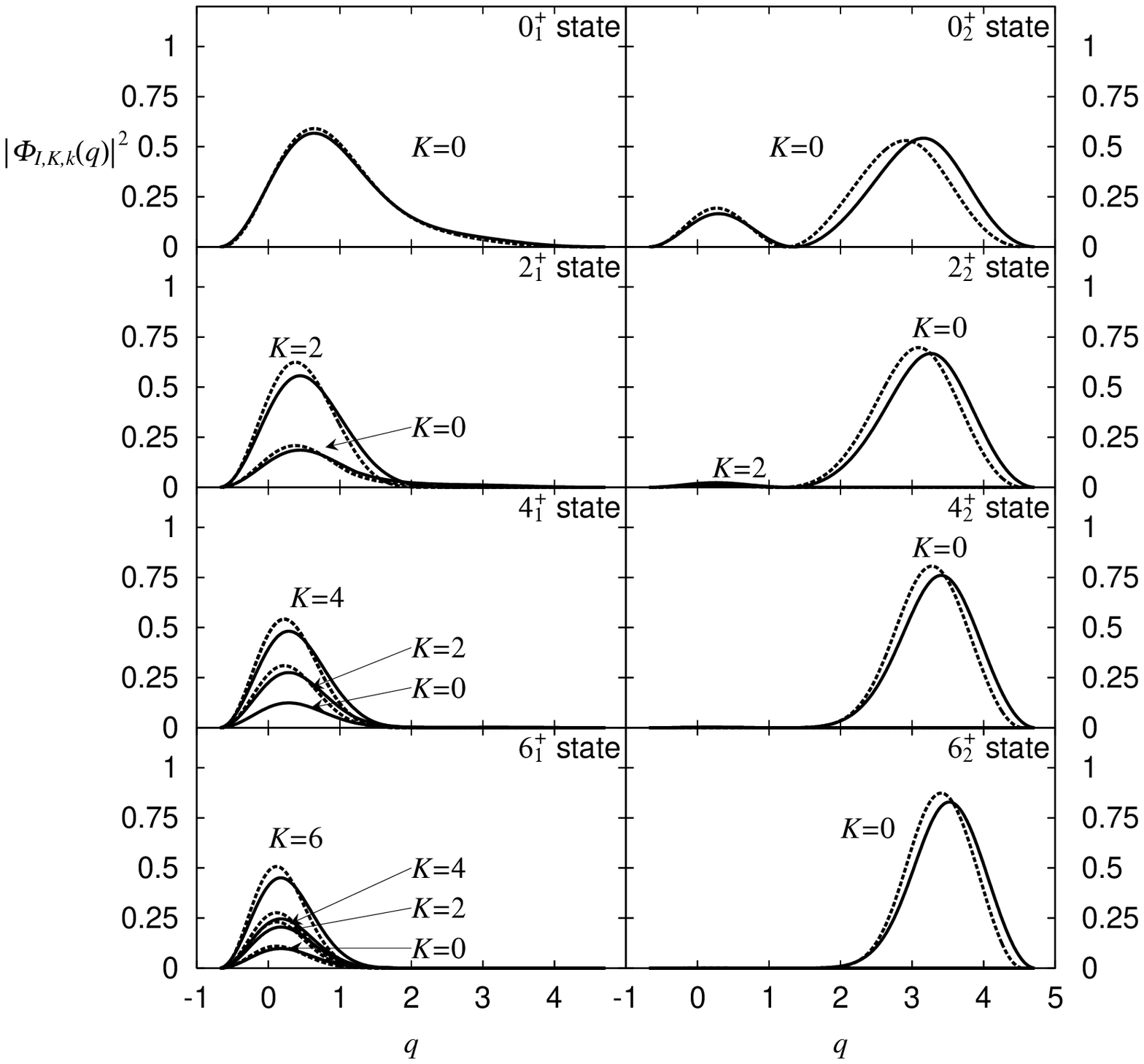}
\end{center}
\caption{Vibrational wave functions squared $|\Phi_{IKk}(q)|^2$
of low-lying states in $^{72}$Kr plotted as functions of $q$.
See the caption to Fig.~\ref{fig:68Se-wavesquared}.}
\label{fig:72Kr-wavesquared}
\end{figure}

\begin{figure}[htbp]
\begin{center}
\begin{tabular}{c}
\includegraphics[width=100mm]{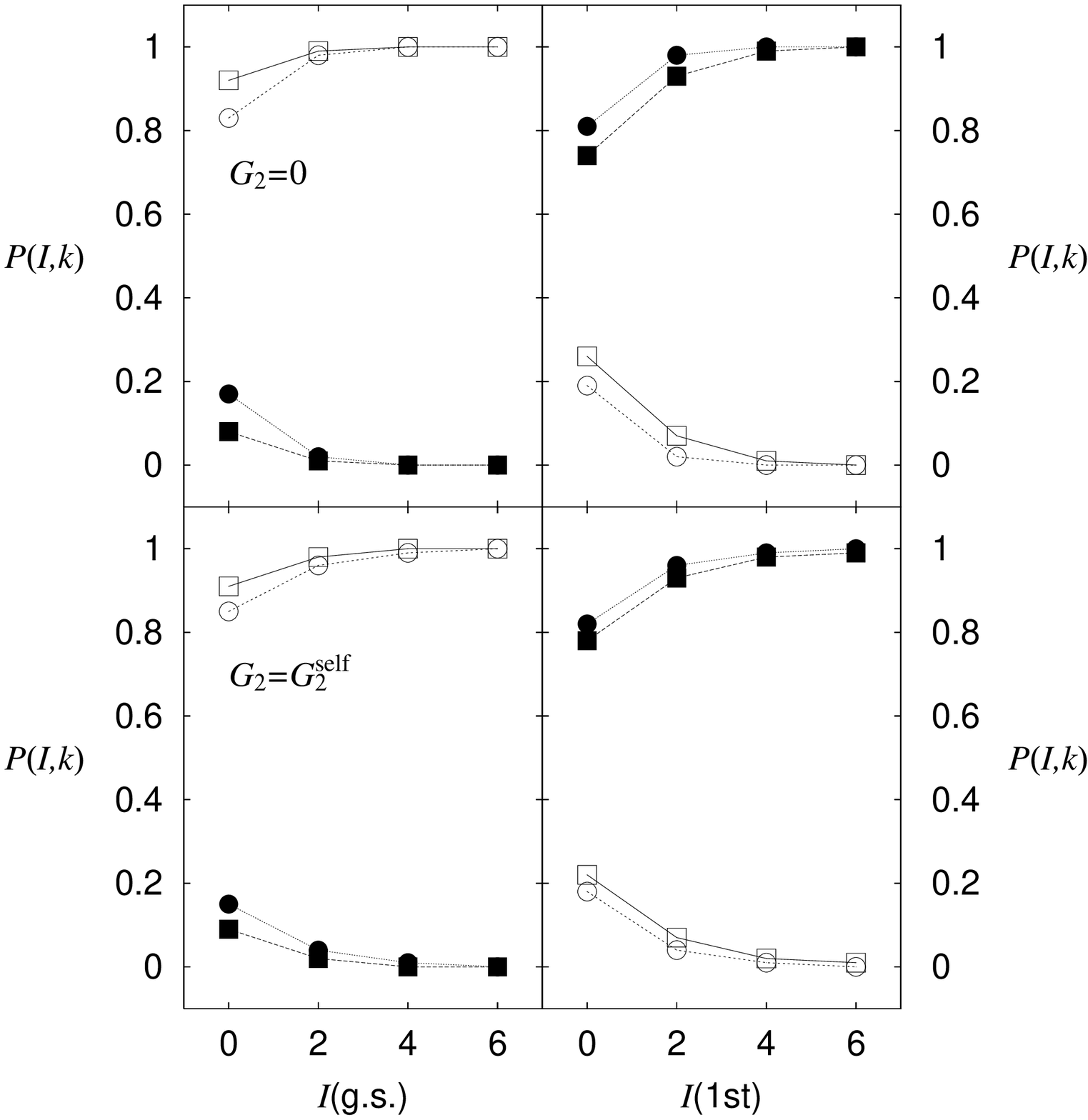}
\end{tabular}
\end{center}
\caption{
The same as Fig.~\ref{fig:68Se-mixing} but for $^{72}$Kr.}
\label{fig:72Kr-mixing}
\end{figure}

For $^{72}$Kr, the collective Schr\"odinger equation is solved
under the boundary conditions (\ref{eq:box-boundary}).
The result of calculation exhibits two coexisting 
rotational bands; see the energy spectra and the $B$(E2) values 
displayed in Fig.~\ref{fig:72Kr-energy}. 
The spectroscopic quadrupole moments presented in Fig.~\ref{fig:72Kr-Qsp} 
indicate that the yrast band possesses the oblate character, while 
the excited band has the prolate character.
For all states including the $0^+$ states, 
the inter-band $B$(E2) values are smaller about one-order of magnitude 
than the intra-band $B$(E2) values, and they rapidly decrease 
as the angular momentum increases.
This indicates that the oblate--prolate shape mixing is rather weak in 
$^{72}$Kr.

In Fig.~\ref{fig:72Kr-wave}, the vibrational wave functions are plotted.
It is seen that the wave function of the $0_1^+$ state is well localized 
in the oblate region, while that of the $0_2^+$ state exhibits the 
major peak in the prolate region.
In the yrast states with $I \ne 0$, the localization about the oblate shape 
further develops for all $K$-components of the vibrational wave functions.
The magnitude is larger for higher $K$. 
In contrast, the collective wave functions of the second lowest states 
in each angular momentum  are essentially composed of the $K=0$ component 
which localize in the prolate region. 
Figure~\ref{fig:72Kr-wavesquared} shows the vibrational wave function
squared. Rather weak oblate--prolate shape mixing is seen only 
for the excited $0^+$ state, and other members of the rotational 
bands possess well--defined oblate or prolate characters.
%The oblate wave functions are composed of all $K$ components,  
%whereas the prolate wave functions are essentially composed of 
%the $K=0$ components.
The oblate and prolate probabilities are presented 
in Fig.~\ref{fig:72Kr-mixing}.
The shape mixing in the $0^+$ states is much weaker compared to $^{68}$Se, 
and almost diminishes at finite angular momentum.

\section{Concluding Remarks} \label{sec:conclusion}

The shape coexistence/mixing phenomena in the low-lying states of $^{68}$Se
and $^{72}$Kr have been investigated using the ASCC method.
The excitation spectra, the spectroscopic quadrupole moments 
and the E2 transition properties of the low-lying states have been 
evaluated for the first time using the ASCC method.
We have derived the quantum collective Hamiltonian 
that describes the coupled collective motion of the large-amplitude 
vibration responsible for the oblate--prolate shape mixing and the 
three-dimensional rotation of the triaxial shape.
The calculation has yielded the excited prolate rotational band
as well as the oblate ground--state band.
The basic pattern of the shape coexistence/mixing phenomena 
has been well reproduced using the one-dimensional collective path
running on the two-dimensional ($\beta, \gamma$) plane. 
This collective path is self-consistently extracted from the huge 
dimensional TDHB manifold. Thus, the result of calculation indicates that 
the TDHB collective dynamics of the shape coexistence/mixing phenomena 
in these nuclei is essentially controlled by the single collective coordinate 
microscopically derived by means of the ASCC method.

We have also shown that the low-lying states are significantly better 
described by including the quadrupole pairing interaction. 
The reason is that the time--odd component of the mean-field generated 
by the quadrupole pairing interaction 
enhances the collective mass of the vibrational motion and the 
moments of inertia of the rotational motion, lowering the energy 
of the collective excitation.

The present calculation clearly indicates that the oblate--prolate shape mixing 
decreases as the angular momentum increases. This implies that 
the rotational dynamics plays the major role to realize   
the localization of vibrational wave functions around the oblate and prolate 
minima under the situation that the barrier between these local minima is 
very low. We shall attempt a more detailed investigation on the dynamical 
reason why the rotational motion hinders the oblate--prolate shape mixing 
in a separate paper. 

\section*{Acknowledgements}

We thank Professors Aiba and Mizutori for useful discussions.
This work is supported by Grants-in-Aid for Scientific Research
(Nos. 18$\cdot$2670, 16540249, 17540231, and 17540244) from the
Japan Society for the Promotion of Science.

\appendix

\section{Quasiparticle Representation of One-Body Operators}
\label{app:quasiparticle}
Since the moving mean-field $\ket{\phi(q)}$ has the positive signature,
the conditions
\begin{align}
 U_{k\mu} = U_{\bar{k}\bar{\mu}}, \quad V_{k\bar{\mu}} = - V_{\bar{k}\mu},
\end{align}
hold.
The matrix elements of the pairing one-body operators 
$\Fhat^{(\pm)}_{s=1,2,3,6,8,11}$ with $K=0,2$ and $r=+1$ 
($\Ahat^{(\tau)(\pm)}, \Bhat_{20(+)}^{(\tau)(\pm)}$, and 
$\Bhat_{22(+)}^{(\tau)(\pm)}$) 
in Eq.~(\ref{eq:F-r=+1,K=0,2}) are 
\begin{align}
 \bra{\phi(q)}\Fhat^{(+)}_{s}\ket{\phi(q)}
=&
 2 {\sum_{\bar{k}l}}' (0|F_{s}^{(+)}|l\bar{k})
 {\sum_{\bar{\mu}}}'
U_{\bar{k}\bar{\mu}}(q) V_{l\bar{\mu}}(q), \\
 F^{(\pm)}_{A,s}(\mu\bar{\nu}) =&
 {\sum_{k\bar{l}}}' (0|F_{s}^{(\pm)}|k\bar{l})
(V_{k\bar{\nu}}(q) V_{\bar{l}\mu}(q) \pm U_{k\mu}(q)
 U_{\bar{l}\bar{\nu}}(q)), \\
 F^{(\pm)}_{B,s}(\mu\nu) =&
 {\sum_{k\bar{l}}}' (0|F_{s}^{(\pm)}|k\bar{l})
(V_{\bar{l}\mu}(q)U_{k\nu}(q) \pm U_{k\mu}(q) V_{\bar{l}\nu}(q)).
\end{align}
The matrix elements of the particle-hole operators with $K=0,2$ and $r=+1$ 
($\Dhat_{20}^{(+)}$ and $\Dhat_{22}^{(+)}$) in Eq.~(\ref{eq:F-r=+1,K=0,2}) are 
\begin{align}
 \bra{\phi(q)}\Fhat^{(+)}_{s=13,15} \ket{\phi(q)} =&
 2 {\sum_{kl}}' (k|F_{s=13,15}^{(+)}|l) {\sum_{\mu}}' V_{k\bar{\mu}}(q)
 V_{l\bar{\mu}}(q), \\
 \Fp_{A,s=13,15}(\mu\bar{\nu}) =&
 {\sum_{kl}}' (k|F_{s=13,15}^{(+)}|l)
(U_{k\mu}(q) V_{l\bar{\nu}}(q) - U_{\bar{k}\bar{\nu}}(q)
 V_{\bar{l}\mu}(q)), \\
 \Fp_{B,s=13,15}(\mu\nu) =&
 {\sum_{kl}}' (k|F^{(+)}_{s=13,15}|l)
(U_{k\mu}(q) U_{l\nu}(q) - V_{\bar{k}\nu}(q) V_{\bar{l}\mu}(q)).
\end{align}
The matrix elements of the particle number operators in Eq.~(\ref{eq:N-qp})
are 
\begin{align}
 N^{(\tau)}(q) =& \bra{\phi(q)}\Nhat^{(\tau)}\ket{\phi(q)}
 = 2 {\sum_{k\in\tau}}' {\sum_{\bar{\mu}}}' V_{k\bar{\mu}}(q)^2, \\
 N_{A}^{(\tau)}(\mu\bar{\nu}) =& {\sum_{k\in\tau}}'
( U_{k\mu}(q) V_{k\bar{\nu}}(q) - U_{\bar{k}\bar{\nu}}(q) V_{\bar{k}\mu}(q)),
 \\
 N_B^{(\tau)}(\mu\nu) =& {\sum_{k\in\tau}}'
( U_{k\mu}(q) U_{k\nu}(q) - V_{\bar{k}\nu}(q) V_{\bar{k}\mu}(q)).
\end{align}
The constraint about the $\Qhat(q-\delta q)$ operator (\ref{eq:Q-const})
in the moving-frame HB equation is written as 
\begin{align}
 \bra{\phi(q)}\Qhat(q-\delta q)\ket{\phi(q)} = 
 {\sum_{kl}}'Q_{kl}(q-\delta q) {\sum_{\mu}}' V_{k\bar{\mu}}(q) V_{l\bar{\mu}}(q).
\end{align}

\section{Determination of the $B$-part of $\Qhat(q)$}
\label{app:Bpart}

We show that the $B$-part of the operator $(\Qhat(q))$ can be determined 
through its $A$-part $(\Qhat^{A}(q))$,
which is obtained by solving the moving-frame QRPA equations.
In terms of the quasiparticle operators, $\adag_i$ and $a_i$, 
defined by the Bogoliubov transformation 
 \begin{align}
  \begin{pmatrix} c \\ \cdag
			    \end{pmatrix} =
  \begin{pmatrix} U & V \\ V^\ast & U^\ast \end{pmatrix}
  \begin{pmatrix} a \\ \adag \end{pmatrix}, 
  \label{eq:Bogoliubov}
 \end{align}
the Hermitian operator $\Qhat(q)$ is written as
\begin{align}
 \Qhat(q) =& \sum_{ij} Q_{ij}(q) \cdag_i c_j \nonumber \\
 =& \sum_{ij}\left( Q^A_{ij}(q) \adag_i \adag_j + Q^{A\ast}_{ij}(q) a_j a_i
+ Q^B_{ij}(q) \adag_i a_j \right),
\end{align}
where 
\begin{align}
 Q^A = \Udag Q V, \quad
 Q^B = \Udag Q U - \Vdag Q V.
\end{align}
Thus the matrices, $Q$ and $Q^B$, can be written in terms of $Q^A$ as 
\begin{align}
 Q   =& (\Udag)^{-1} Q^A V^{-1}, \label{eq:QfromQA}\\
 Q^B =& Q^A V^{-1} U - \Vdag (\Udag)^{-1} Q^A. \label{eq:QBfromQA}
\end{align}
We cannot directly use these relations, however, for determining $Q$ 
and $Q^B$, because the matrices $Q$ and $Q^B$ calculated by 
(\ref{eq:QfromQA}) and (\ref{eq:QBfromQA}) are not Hermitian.
We have to construct Hermitian matrices $Q$ and $Q^B$ from $Q^A$.
This is achieved by adding a symmetric matrix $S$ to the
solution of the moving-frame QRPA equation, which we here denote 
$Q_0^A$, as 
\begin{align}
 Q^A = Q_0^A + S.
\end{align}
For this matrix, the operator $Q$ is written as
\begin{align}
 Q =& (\Udag)^{-1} (Q^A_0+ S) V^{-1} = Q_0 +
 (\Udag)^{-1}SV^{-1},\label{eq:StoQ}\\
 \Qdag =& (\Vdag)^{-1} (Q^A_0+S)^\dagger U^{-1} = \Qdag_0 +
 (\Vdag)^{-1}S^\dagger U^{-1}.
\end{align}
From the Hermicity condition, $Q=\Qdag$, we obtain the following equation 
for $S$.  
\begin{align}
 (\Vdag)^{-1} S^\dagger U^{-1} - (\Udag)^{-1}SV^{-1} = Q_0 - \Qdag_0, 
 \label{eq:Q-QT}
\end{align}
which determine the symmetric matrix $S$.  Let us write the above equation 
explicitly, 
\begin{align}
 \sum_{kl} \{
 (V^{-1})_{ki} (U^{-1})_{lj} - (U^{-1})_{ki} (V^{-1})_{lj}
\} S_{kl} =& (Q_0)_{ij} - (Q_0)_{ji},
\label{eq:Sdetermine}
\end{align}
where we assume that all quantities are real.
Both the number of unknown quantity and number of equations are the same, 
$N(N+1)/2$, $N$ being the dimension of the matrix. Therefore
it is possible to determine the matrix $S$ by solving this equation.

In the case of the P+Q model, we start from the skew symmetric matrix $Q^A_0$,
\begin{align}
 (Q^A_0)_{\mu\bar{\nu}}(q) = \frac{1}{2} Q^A_{\mu\bar{\nu}}(q), \quad
 (Q^A_0)_{\bar{\nu}\mu}(q) =-\frac{1}{2} Q^A_{\mu\bar{\nu}}(q).
\end{align}
The symmetric matrix $S_{\mu\bar{\nu}} = S_{\bar{\nu}\mu}$ 
is determined by solving the following equation: 
\begin{align}
{\sum_{\mu\bar{\nu}}}'\{ (V^{-1})_{\bar{\nu}k} (U^{-1})_{\mu l} -
(U^{-1})_{\mu k} (V^{-1})_{\bar{\nu}l} \} S_{\mu\bar{\nu}}
=& (Q_0)_{kl} - (Q_0)_{lk}  \label{eq:Sdetermine-PQQ1} \\
{\sum_{\mu\bar{\nu}}}'\{ (V^{-1})_{\mu\bar{k}}
 (U^{-1})_{\bar{\nu}\bar{l}}  -
 (U^{-1})_{\bar{\nu}\bar{k}} (V^{-1})_{\mu\bar{l}}
\} S_{\mu\bar{\nu}} =& (Q_0)_{\bar{k}\bar{l}} -
 (Q_0)_{\bar{l}\bar{k}} \label{eq:Sdetermine-PQQ2}
\end{align}
where
\begin{align}
 (Q_0)_{kl}(q) &= {\sum_{\mu\bar{\nu}}}' (U^{-1})_{\mu k}
 (Q^A_0)_{\mu\bar{\nu}}(q)
 (V^{-1})_{\bar{\nu}l}, \\
 (Q_0)_{\bar{k}\bar{l}}(q) &= {\sum_{\mu\bar{\nu}}}'
 (U^{-1})_{\bar{\nu}\bar{k}} (Q^A_0)_{\bar{\nu}\mu}(q)
 (V^{-1})_{\mu\bar{l}}.
\end{align}
As the relation $(Q_0^A)_{\mu\bar{\nu}} = (Q_0^A)_{\nu\bar{\mu}}$
holds, the matrix $S$ satisfies the relation $S_{\mu\bar{\nu}} =
-S_{\nu\bar{\mu}}$, 
and then, the Eqs.~(\ref{eq:Sdetermine-PQQ1}) and (\ref{eq:Sdetermine-PQQ2})
are written as
\begin{align}
 {\sum_{\mu\bar{\nu}}}'\{ -(V^{-1})_{\bar{\nu}k} (U^{-1})_{\mu l} -
(U^{-1})_{\nu k} (V^{-1})_{\bar{\mu}l} \} S_{\mu\bar{\nu}}
=& (Q_0)_{kl} - (Q_0)_{lk}.
\end{align}
Using the transformed matrix $Q^{A'}(q)$
\begin{align}
 Q^{A'}_{\mu\bar{\nu}}(q) &= (Q^A_0)_{\mu\bar{\nu}}(q) + S_{\mu\bar{\nu}},
\end{align}
the Hermite matrices $Q(q)$ and $Q^B(q)$ are obtained as follows: 
\begin{align}
 Q_{kl}(q) &= {\sum_{\mu\bar{\nu}}}' (U^{-1})_{\mu
 k}Q^{A'}_{\mu\bar{\nu}}(q)(V^{-1})_{\bar{\nu}l}, \\
 Q_{\bar{k}\bar{l}}(q) &= {\sum_{\bar{\mu}\nu}}'
(U^{-1})_{\bar{\mu} \bar{k}}
Q^{A'}_{\bar{\mu}\nu}(q)(V^{-1})_{\nu\bar{l}},  \\
 Q^B_{\mu\nu}(q) &= {{\sum_{kl}}}' U_{k\mu} Q_{kl}(q) U_{l\nu}
 - V_{\bar{k}\mu} Q_{\bar{k}\bar{l}}(q) V_{\nu\bar{l}}, \\
 Q^B_{\bar{\mu}\bar{\nu}}(q) &= {\sum_{\bar{k}\bar{l}}}'
 U_{\bar{\mu}\bar{k}}Q_{\bar{k}\bar{l}}(q)U_{\bar{l}\bar{\nu}}
-V_{k\bar{\mu}} Q_{kl}(q) V_{l\bar{\nu}}.
\end{align}

\section{Calculation of the Rotational Moments of Inertia}
\label{app:MoI}

For the separable interactions (\ref{eq:separableH}), 
the Thouless--Valatin equations 
(\ref{eq:can-qI}) determining the three rotational moments of inertia 
$\Jc_i(q)$ about the principal axes 
at a non-equilibrium state $\ket{\phi(q)}$ 
can be written in the following form: 
 \begin{multline}
\delta \bra{\phi(q)}[\hhat_M(q),\Psihat_i(q)] + i\sum_s
 \fp_{\Psi_i,s}(q)\Fhatp_s - \sum_s \fm_{\Psi_i,s}(q) \Fhatm_s \\
 - \frac{1}{i}\Jc_i^{-1}(q) \Ihat_i \ket{\phi(q)} = 0,
  \label{eq:asccrot_sep}
\end{multline}
where
\begin{align}
  \fp_{\Psi_i,s}(q) &= i\kappa_s \bra{\phi(q)}[\Fhatp_s,
 \Psihat_i(q)]\ket{\phi(q)}, \\
 \fm_{\Psi_i,s}(q) &=-\kappa_s \bra{\phi(q)}[\Fhatm_s,
 \Psihat_i(q)]\ket{\phi(q)}.
\end{align}
The quasiparticle representation of the angular momentum operators are
\begin{align}
 \Ihat_x =& {\sum_{\mu\nu}}' I_{A,x}(\mu\bar{\nu})(\Abdag_{\mu\bar{\nu}} +
 \Ab_{\mu\bar{\nu}}) + {\sum_{\mu\nu}}' I_{B,x}(\mu\nu)(\Bb_{\mu\nu} -
 \Bb_{\bar{\mu}\bar{\nu}}), \label{eq:Jx-qp} \\
 i\Ihat_y =& {\sum_{\mu\nu}}' I_{A,y}(\mu\nu) (\Abdag_{\mu\nu} -
 \Ab_{\mu\nu})
 + I_{A,y}(\bar{\mu}\bar{\nu})
 (\Abdag_{\bar{\mu}\bar{\nu}} - \Ab_{\bar{\mu}\bar{\nu}})
  \nonumber \\
 &+ {\sum_{\mu\nu}}' I_{B,y}(\mu\bar{\nu})( \Bb_{\mu\bar{\nu}} -
 \Bb_{\bar{\mu}\nu}),
  \label{eq:Jy-qp} \\
 \Ihat_z =& {\sum_{\mu\nu}}' I_{A,z}(\mu\nu) (\Abdag_{\mu\nu} +
 \Ab_{\mu\nu})
 + I_{A,z}(\bar{\mu}\bar{\nu})
 (\Abdag_{\bar{\mu}\bar{\nu}} + \Ab_{\bar{\mu}\bar{\nu}}) \nonumber \\
 &+ {\sum_{\mu\nu}}' I_{B,z}(\mu\bar{\nu})( \Bb_{\mu\bar{\nu}} +
 \Bb_{\bar{\mu}\nu}), \label{eq:Jz-qp}
\end{align}
where the matrix elements of $I_{A,x}, I_{A,y}$ and $I_{A,z}$ are 
given by 
\begin{align}
 I_{A,x}(\mu\bar{\nu}) =& {\sum_{kl}}' (k|I_x|l)
(U_{k\mu}(q) V_{l\bar{\nu}}(q)  + U_{\bar{k}\bar{\nu}}(q)
 V_{\bar{l}\mu}(q)), \\
 I_{A,y}(\mu\nu) =& I_{A,y}(\bar{\mu}\bar{\nu}) = {\sum_{kl}}' (k|I_y|l)
 U_{k\mu}(q) V_{\bar{l}\nu}(q), \\
 I_{A,z}(\mu\nu) =& - I_{A,z}(\bar{\mu}\bar{\nu}) =
 - {\sum_{k}}' m_k U_{k\mu}(q) V_{\bar{k}\nu}(q).
\end{align}
The residual interactions with ($r=+1,K=1$), ($r=-1,K=1$) and ($r=-1$,$K=2$) 
contribute to rotations about the $x$, $y$ and $z$-axis, respectively.
The quasiparticle representation of one-body operators having these 
quantum numbers are 
\begin{align}
 \Fhat^{(\pm)}_s =& {\sum_{\mu\bar{\nu}}}' F^{(\pm)}_{A,s}(\mu\bar{\nu})
(\Abdag_{\mu\bar{\nu}} \pm \Ab_{\mu\bar{\nu}}) \nonumber \\
&+ {\sum_{\mu\nu}}' F^{(\pm)}_{B,s}(\mu\nu)
(\Bb_{\mu\nu} - \Bb_{\bar{\mu}\bar{\nu}}), \quad (r=+1, K=1) \\
 \Fhat^{(\pm)}_s =& {\sum_{\mu\nu}}' F^{(\pm)}_{A,s}(\mu\nu)
(\Abdag_{\mu\nu} \pm \Ab_{\mu\nu}
+ \Abdag_{\bar{\mu}\bar{\nu}} \pm \Ab_{\bar{\mu}\bar{\nu}}) \nonumber \\
&+ {\sum_{\mu\nu}}' F^{(\pm)}_{B,s}(\mu\bar{\nu}) (\Bb_{\mu\bar{\nu}}
 -\Bb_{\bar{\mu}\nu}), \quad (r = -1, K=1) \\
 \Fhat^{(\pm)}_s =& {\sum_{\mu\nu}}' F^{(\pm)}_{A,s}(\mu\nu)
\{\Abdag_{\mu\nu} \pm \Ab_{\mu\nu}
- (\Abdag_{\bar{\mu}\bar{\nu}} \pm \Ab_{\bar{\mu}\bar{\nu}})\} \nonumber
 \\
&+ {\sum_{\mu\nu}}' F^{(\pm)}_{B,s}(\mu\bar{\nu}) (\Bb_{\mu\bar{\nu}}
 +\Bb_{\bar{\mu}\nu}),\quad  (r = -1, K=2).
\end{align}
The matrix elements of the quadrupole pairing operators are
\begin{align}
 F^{(\pm)}_{A,s=4,9}(\mu\bar{\nu}) =& 2 {\sum_{kl\in\tau}}'
 (0|B^{(\tau)(\pm)}_{21(-)}|k\bar{l})
( V_{\bar{l}\mu}(q) V_{k\bar{\nu}}(q) \pm U_{k\mu}(q)
 U_{\bar{l}\bar{\nu}}(q)), \\
 F^{(\pm)}_{A,s=5,10}(\mu\nu) =&
 {\sum_{kl\in\tau}}'  (0|B^{(\tau)(\pm)}_{21(+)}|kl)
 (V_{l\bar{\mu}}(q) V_{k\bar{\nu}}(q) \pm U_{\bar{k}\bar{\mu}}(q)
 V_{\bar{l}\bar{\nu}}(q)), \\
 F^{(\pm)}_{A,s=7,12}(\mu\nu) =&
 {\sum_{kl\in\tau}}' (0|B^{(\tau)(\pm)}_{22(-)}|kl)
 (- V_{\bar{l}\mu}(q) V_{\bar{k}\nu}(q) \pm U_{k\mu}(q)
 U_{l\nu}(q)), \\
 F^{(\pm)}_{A,s=5,10}(\mu\nu) =&
 F^{(\pm)}_{A,s=5,10}(\bar{\mu}\bar{\nu}),  \quad
 F^{(\pm)}_{A,s=7,12}(\mu\nu) =
-F^{(\pm)}_{A,a=7,12}(\bar{\mu}\bar{\nu}).
\end{align}
Note that the quadrupole operators $\Dhat_{21}^{(\pm)},
\Dhat_{22}^{(-)}$ does not contribute to the moments of inertia.

The three Thouless--Valatin equations, (\ref{eq:asccrot_sep}), 
at non-equilibrium can be solved independently.
The angle operators, $\Psihat_x(q), \Psihat_y(q)$ and $\Psihat_z(q)$,
can be written as 
\begin{subequations}
\begin{align}
\Psihat_x(q) &= i{\sum_{\mu\bar{\nu}}}' \Psi_{A,x}(\mu\bar{\nu})
(\Abdag_{\mu\bar{\nu}} - \Ab_{\mu\bar{\nu}}) + (\Bb-{\rm part}),
\\
\Psihat_y(q) &= {\sum_{\mu\nu}}' \Psi_{A,y}(\mu\nu)( \Abdag_{\mu\nu} +
 \Ab_{\mu\nu}) +
 \Psi_{A,y}(\bar{\mu}\bar{\nu}) (\Abdag_{\bar{\mu}\bar{\nu}} +
 \Ab_{\bar{\mu}\bar{\nu}})
 + (\Bb {\rm -part}), \\
\Psihat_z(q) &= i {\sum_{\mu\nu}}' \Psi_{A,z}(\mu\nu) (\Abdag_{\mu\nu}
 -\Ab_{\mu\nu})
 +  \Psi_{A,z}(\bar{\mu}\bar{\nu})
 (\Abdag_{\bar{\mu}\bar{\nu}}
 - \Ab_{\bar{\mu}\bar{\nu}}) + (\Bb {\rm -part}). 
\end{align} \label{eq:Psi_ME}
\end{subequations} 
These matrix elements are easily obtained from Eq.~(\ref{eq:asccrot_sep}) as 
\begin{subequations}
\begin{align}
 \Psi_{A,x}(\mu\bar{\nu}) &= \frac{-1}{E_\mu + E_{\bar{\nu}}}
 \left(
 \sum_s \fp_{\Psi_x,s}(q)\Fp_{A,s}(\mu\bar{\nu}) + \Jc^{-1}_x(q)
 I_{A,x}(\mu\bar{\nu}) \right), \\
 \Psi_{A,y}(\mu\nu) &= \frac{1}{E_\mu + E_{\nu}}
\left(
  \sum_s \fm_{\Psi_y,s}(q)\Fm_{A,s}(\mu\nu) - \Jc^{-1}_y(q)
 I_{A,y}(\mu\nu)
\right), \\
 \Psi_{A,y}(\bar{\mu}\bar{\nu}) &= \frac{1}{E_{\bar{\mu}} +
 E_{\bar{\nu}}}
\left(
  \sum_s \fm_{\Psi_y,s}(q)\Fm_{A,s}(\bar{\mu}\bar{\nu}) - \Jc^{-1}_y(q)
 I_{A,y}(\bar{\mu}\bar{\nu})
\right), \\
  \Psi_{A,z}(\mu\nu) &= \frac{1}{E_{\mu} + E_{\nu}}
 \left(
 - \sum_s \fp_{\Psi_z,s}(q)\Fp_{A,s}(\mu\nu) - \Jc_z^{-1}(q)
 I_{A,z}(\mu\nu)
\right), \\
 \Psi_{A,z}(\bar{\mu}\bar{\nu}) &= \frac{1}{E_{\bar{\mu}} +
 E_{\bar{\nu}}}
 \left(
 - \sum_s \fp_{\Psi_z,s}(q)\Fp_{A,s}(\bar{\mu}\bar{\nu}) - \Jc_z^{-1}(q)
 I_{A,z}(\bar{\mu}\bar{\nu})
\right).
\end{align} \label{eq:Psi_ME2}
\end{subequations}
It is easy to confirm that 
$\fm_{\Psi_x,s}(q)=0, \fp_{\Psi_y,s}(q) = 0$ and $\fm_{\Psi_z,s}(q) = 0$.
Substituting (\ref{eq:Psi_ME2}) into the quantities, 
$\fp_{\Psi_x,s}(q), \fm_{\Psi_y,s}(q), \fp_{\Psi_z,s}(q)$,  
and the canonical variable condition (\ref{eq:can-qI}), we obtain 
\begin{subequations}
\label{eq:fPsi}
\begin{align}
 \fp_{\Psi_x,s}(q) &= i \kappa_s \bra{\phi(q)}[\Fhatp_s,
 \Psihat_x(q)]\ket{\phi(q)}
 \nonumber  \\
 &= 2\kappa_s \sum_{s'} (\Fp_{A,s}, \Fp_{A,s'})_{E+} \fp_{\Psi_x,s'}(q)
 +
 2\kappa_s (\Fp_{A,s}, I_{A,x})_{E+} \Jc_x^{-1}(q), \label{eq:fPsi_x} \\
\fm_{\Psi_y,s}(q) =& - \kappa_s \bra{\phi(q)}[\Fhatm_s, \Psihat_y]
 \ket{\phi(q)}
\nonumber \\
=& 2\kappa_s \sum_{s'}  (\Fm_{A,s}, \overline{\Fm_{A,s'}})_{E-}
 \fm_{\Psi_y, s'}(q)
- 2\kappa_s(\Fm_{A,s},\overline{I_{A,y}})_{E-}\Jc^{-1}_y(q),
 \label{eq:fPsi_y} \\
 \fp_{\Psi_z,s}(q) =& i \kappa_s \bra{\phi(q)}[\Fhatp_s,
 \Psihat_z(q)]\ket{\phi(q)} \nonumber \\
=& 2\kappa_s \sum_{s'} (\Fp_{A,s},\overline{I_{A,z}})_{E-}
 \fp_{\Psi_z,s'}(q)
 + 2\kappa_s (\Fp_{A,s}, \overline{I_{A,z}})_{E-} \Jc_z^{-1}(q),
\end{align}
\end{subequations}
\begin{subequations}
\label{eq:can-qI-xyz}
\begin{align}
 \bra{\phi(q)}[\Psihat_x(q), \Ihat_x]\ket{\phi(q)}/i &=
 -2 {\sum_{\mu\bar{\nu}}}' \Psi_{A,x}(\mu,\bar{\nu}) I_{A,x}(\mu\bar{\nu})
 \nonumber \\
&=  -2 \sum_s (\Fp_{A,s}, I_{A,x})_{E+} \fp_{\Psi_x,s}(q) -2
(I_{A,x},I_{A,x})_{E+} \Jc_x^{-1}(q) \nonumber \\
&= 1, \label{eq:can-qI-x}
 \\
\bra{\phi(q)}[\Psihat_y(q),i\Ihat_y]\ket{\phi(q)}
 &= 2{\sum_{\mu\nu}}' \Psi_{A,y}(\mu\nu) \overline{I_{A,y}}(\mu\nu)
+ 2{\sum_{\bar{\mu}\bar{\nu}}}' \Psi_{A,y}(\bar{\mu}\bar{\nu})
 \overline{I_{A,y}}(\bar{\mu}\bar{\nu}) \nonumber \\
 &= 2\sum_s (\Fm_{A,s}, \overline{I_{A,y}})_{E-} \fm_{\Psi_y,s}(q)
 - 2 (I_{A,y}, \overline{I_{A,y}})_{E-} \Jc_y^{-1}(q) \nonumber \\
 &= -1,
 \label{eq:can-qI-y}
\\
 \bra{\phi(q)}[\Psihat_z(q), \Ihat_z]\ket{\phi(q)}/i &=
 -2{\sum_{\mu\nu}}' \Psi_{A,z}(\mu\nu)\overline{I_{A,z}}(\mu\nu)
 -2{\sum_{\bar{\mu}\bar{\nu}}}' \Psi_{A,z}(\bar{\mu}\bar{\nu})
 \overline{I_{A,z}}(\bar{\mu}\bar{\nu}) \nonumber \\
 &= 2\sum_s (\Fp_{A,s},\overline{I_{A,z}})_{E-} \fp_{\Psi_z,s'}(q)
 + 2(I_{A,z}, \overline{I_{A,z}})_{E-} \Jc_z^{-1}(q) \nonumber \\
 &= 1, \label{eq:can-qI-z}
\end{align}
\end{subequations}
where
\begin{align}
 (X, Y)_{E+} &= {\sum_{\mu\bar{\nu}}}'
 \frac{X(\mu\bar{\nu}) Y(\mu\bar{\nu})}{E_{\mu} + E_{\bar{\nu}}}, \\
 \overline{X}(\mu\nu) &= X(\mu\nu) - X(\nu\mu), \\
 (X,Y)_{E-} &= {\sum_{\mu\nu}}' \frac{X(\mu\nu) Y(\mu\nu)}{E_{\mu} +
 E_{\nu}}
  + {\sum_{\bar{\mu}\bar{\nu}}}' \frac{X(\bar{\mu}\bar{\nu})
 Y(\bar{\mu}\bar{\nu})}{E_{\bar{\mu}} + E_{\bar{\nu}}}.
\end{align}
Equations (\ref{eq:fPsi}) and (\ref{eq:can-qI-xyz}) are linear
equations with respect to $\fp_{\Psi_i,s}(q)$ and $\Jc_i^{-1}(q)$,
and can be rewritten as follows:
\begin{subequations}
\begin{align}
\sum_{s'}
\begin{pmatrix}
2\kappa_s (\Fp_{A,s},\Fp_{A,s'})_{E+} - \delta_{ss'} & 2\kappa_{s'}
 (\Fp_{A,s}, I_{A,x})_{E+} \\
 2(\Fp_{A,s}, I_{A,x})_{E+} & 2(I_{A,x}, I_{A,x})_{E+}, \\
\end{pmatrix}
\begin{pmatrix}
 \fp_{\Psi_x,s'}(q) \\ \Jc_x^{-1}(q)
\end{pmatrix}
= \begin{pmatrix} 0 \\ 1 \end{pmatrix},
\\
 \sum_{s'}
\begin{pmatrix}
2\kappa_s (\Fm_{A,s}, \overline{\Fm_{A,s'}})_{E-}-\delta_{ss'} &
 -2\kappa_s
 (\Fm_{A,s},\overline{I_{A,y}})_{E-} \\
 -2 (\Fm_{A,s}, \overline{I_{A,y}})_{E-} & 2
 (I_{A,y},\overline{I_{A,y}})_{E-}
\end{pmatrix}
\begin{pmatrix}
 \fm_{\Psi_y,s}(q) \\ \Jc_y^{-1}
\end{pmatrix}
=
\begin{pmatrix}
0 \\ 1
\end{pmatrix},
\\
\sum_{s'}
\begin{pmatrix}
 2\kappa_s(\Fp_{A,s}, \overline{\Fp_{A,s}})_{E-} - \delta_{ss'} &
 2\kappa_s
 (\Fp_{A,s}, \overline{I_{A,z}})_{E-} \\
 2 (\Fp_{A,s'}, \overline{I_{A,z}})_{E-} & 2 (I_{A,z},
 \overline{I_{A,z}})_{E-}
\end{pmatrix}
\begin{pmatrix}
 \fp_{\Psi_z, s'}(q) \\ \Jc^{-1}_z(q)
\end{pmatrix}
=
\begin{pmatrix}
 0 \\ 1
\end{pmatrix}.
\end{align}
\end{subequations}

\end{document}